\documentclass[twocolumn]{aastex631}

\usepackage{xcolor}
\usepackage{subfloat}
\usepackage{multirow}
\usepackage{soul}
\usepackage{graphicx}	
\usepackage{amsmath}	

\usepackage{amssymb}	
\usepackage{cleveref}   
\usepackage{ulem}
\usepackage{multirow}
\usepackage{longtable}
\usepackage{comment}

\begin{document}

\title{The Transit Timing and Transmission Spectrum of Hot Jupiter WASP-43~b from a decade of Multi-band Transit Follow-up Observations}

\author[0000-0001-7234-7167]{Napaporn A-thano}
\affiliation{National Astronomical Research Institute of Thailand, 260 Moo 4, Donkaew, Mae Rim, Chiang Mai, 50180, Thailand}
\email{napaporn@narit.or.th}

\author[0000-0003-3251-3583]{Supachai Awiphan}
\affiliation{National Astronomical Research Institute of Thailand, 260 Moo 4, Donkaew, Mae Rim, Chiang Mai, 50180, Thailand}
\email{supachai@narit.or.th}

\author[0000-0002-1743-4468]{Eamonn Kerins}
\affiliation{Jodrell Bank Centre for Astrophysics, University of Manchester, Oxford Road, Manchester, M13 9PL, UK}

\author[0000-0003-1143-0877]{Akshay Priyadarshi}
\affiliation{Jodrell Bank Centre for Astrophysics, University of Manchester, Oxford Road, Manchester, M13 9PL, UK}

\author[0000-0003-0356-0655]{Iain McDonald}
\affiliation{Jodrell Bank Centre for Astrophysics, University of Manchester, Oxford Road, Manchester, M13 9PL, UK}
\affiliation{Department of Physical Sciences, The Open University, Walton Hall, Milton Keynes, MK7 6AA, UK}

\author[0000-0001-7359-3300]{Ing-Guey Jiang}
\affiliation{Department of Physics and Institute of Astronomy, National Tsing-Hua University, Hsinchu 30013, Taiwan}

\author[0000-0001-8657-1573]{Yogesh C. Joshi}
\affiliation{Aryabhatta Research Institute of Observational Sciences (ARIES), Manora Peak, Nainital 263001, India}

\author[0000-0002-6039-8212]{Fan Yang}
\affil{D\'epartement d'Astrophysique/AIM, CEA/IRFU, CNRS/INSU, Univ. Paris-Saclay, Univ. de Paris, 91191 Gif-sur-Yvette, France\\}

\author{Ida Janiak}
\affiliation{Jodrell Bank Centre for Astrophysics, University of Manchester, Oxford Road, Manchester, M13 9PL, UK}

\author{Patcharawee Munsaket}
\affiliation{School of Physics, Institute of Science, Suranaree University of Technology, 111 University Ave., \\ Suranaree, Nakhon Ratchasima 30000, Thailand}

\author[0000-0001-5162-4225]{Yasir Abdul Qadir}
\affil{Department of Physics and Astronomy, FI-20014 University of Turku, Finland\\}

\author{Ronnakrit Rattanamala}
\affiliation{Department of Physics and General Science, Faculty of Science and Technology, \\ Nakhon Ratchasima Rajabhat University, Nakhon Ratchasima, 30000, Thailand}

\author{Orarik Tasuya}
\affiliation{National Astronomical Research Institute of Thailand, 260 Moo 4, Donkaew, Mae Rim, Chiang Mai, 50180, Thailand}
\affiliation{Department of Physics and Materials Science, Faculty of Science, Chiang Mai University, Chiang Mai 50200, Thailand}

\author{Ekburus Boonsoy}
\affiliation{School of Physics, Institute of Science, Suranaree University of Technology, 111 University Ave., \\ Suranaree, Nakhon Ratchasima 30000, Thailand}
\affiliation{Center of Excellence for Entrepreneurship, Suranaree University of Technology, Nakhon Ratchasima, 30000 Thailand}

\author{Nuanwan Sanguansak}
\affiliation{School of Physics, Institute of Science, Suranaree University of Technology, 111 University Ave., \\ Suranaree, Nakhon Ratchasima 30000, Thailand}
\affiliation{National Science and Technology Development Agency, Thailand Science Park, Pathum Thani 12120, Thailand)}

\author{Rattiyakorn Rattanasai}
\affiliation{National Astronomical Research Institute of Thailand, 260 Moo 4, Donkaew, Mae Rim, Chiang Mai, 50180, Thailand}
\affiliation{Department of Physics and Materials Science, Faculty of Science, Chiang Mai University, Chiang Mai, 50200, Thailand}

\author{Thammasorn Padjaroen}
\affiliation{Department of Physics and Materials Science, Faculty of Science, Chiang Mai University, Chiang Mai, 50200, Thailand}

\author{Siramas Komonjinda}
\affiliation{Department of Physics and Materials Science, Faculty of Science, Chiang Mai University, Chiang Mai, 50200, Thailand}

\author{Sawatkamol Pichadee}
\affiliation{National Astronomical Research Institute of Thailand, 260 Moo 4, Donkaew, Mae Rim, Chiang Mai, 50180, Thailand}

\author{Ananpol Sudsap}
\affiliation{National Astronomical Research Institute of Thailand, 260 Moo 4, Donkaew, Mae Rim, Chiang Mai, 50180, Thailand}

\author{Smanchan Chandaiam}
\affiliation{National Astronomical Research Institute of Thailand, 260 Moo 4, Donkaew, Mae Rim, Chiang Mai, 50180, Thailand}

\author{Boonyarit Choonhakit}
\affiliation{National Astronomical Research Institute of Thailand, 260 Moo 4, Donkaew, Mae Rim, Chiang Mai, 50180, Thailand}

\author{Suwanit Wutsang}
\affiliation{National Astronomical Research Institute of Thailand, 260 Moo 4, Donkaew, Mae Rim, Chiang Mai, 50180, Thailand}

\author{Vik S Dhillon}
\affiliation{Department of Physics and Astronomy, University of Sheffield, Sheffield, S3 7RH, UK}
\affiliation{Instituto de Astrofísica de Canarias, E-38205 La Laguna, Tenerife, Spain}



\begin{abstract}

We present a new set of 35 transit light curves of the hot Jupiter WASP-43~b, obtained through the SPEARNET network. These datasets were analyzed together with previously published ground-based observations, as well as space-based data from \emph{TESS}, \emph{HST}, and \emph{JWST}, to refine the planetary parameters of WASP-43~b. A total of 188 mid-transit times, measured with \texttt{TransitFit}, were analyzed for potential timing variations. The transit timing variations do not show any significant evidence of orbital decay. Atmospheric retrievals using \emph{HST}/WFC3 G141 transmission spectra suggest that higher-temperature solutions are associated with higher water abundances. However, when these data are combined with observations from ground-based telescopes, \emph{TESS}, and \emph{JWST}, the increased modeling complexity across the broad wavelength baseline presents significant challenges for atmospheric characterization. These results highlight that high-precision, multi-instrument datasets will be necessary to break existing degeneracies in the atmospheric modeling of this target in the future.

\end{abstract}



\section{Introduction}

WASP-43b, a hot Jupiter with an ultrashort orbital period of 0.81 days around the young K7 dwarf star WASP-43 ($V$=12.4), was initially discovered by the Wide-Angle Search for Planets survey (WASP), \citet{hellier2011}. WASP-43~b is a prime candidate for investigating Transit Timing Variations (TTVs) due to its short orbital period. The first TTV study was conducted by \citet{gillon2012}, who found that the transit timing residuals diagram ($O-C$, observed minus computed) indicated that the mid-transit times were consistent with a linear ephemeris. \citet{maciejewski2013} found that the orbital period of WASP-43~b was shorter by 0.13 seconds compared to the value reported by \citet{gillon2012}. In 2014, \citet{blecic2014,murgas2014} applied a quadratic fitting model to the transit-timing data and reported a potential orbital period decay. However, studies by \citet{chen2014,ricci2015} noted that there was no significant evidence for orbital decay in WASP-43~b. In 2016, \citet{jiang2016} presented a possible orbital decay in WASP-43b with a rate of $\dot{P} = -0.029 \pm 0.008$ sec/year. However, \citet{hoyer2016} ruled out the orbital decay of WASP-43b, reporting a period change rate of $\dot{P} = -0.02 \pm 6.6$ ms/year, which is three orders of magnitude smaller than the value previously reported by \citet{jiang2016}. \citet{stevenson2017} also found no evidence for tidal decay. \citet{patra2020} found a positive orbital change rate of $\dot{P} = (1.9 \pm 0.6) \times 10^{-10}$.

After the launch of the \emph{Transiting Exoplanet Survey Satellite} (\emph{TESS}), \citet{wong2020} updated the transit ephemerides, providing an orbital period change of $\dot{P} < 5.6$ ms/year. Recently, \citet{davoudi2021} found a decrease in the orbital period of WASP-43b with a rate of $-0.0035 \pm 0.0007$ seconds/year. Moreover, \citet{garai2021} conducted another study using \emph{TESS} data combined with data from the Multicolor Simultaneous Camera for studying Atmospheres of Transiting exoplanets2 (MuSCAT2) and found the orbital period change rate of WASP-43b to be $\dot{P} = -0.6 \pm 1.2$ ms/year, which is consistent with a constant period value. Therefore, they confirmed that no orbital decay was detected in WASP-43b. Based on these developments, further investigation is needed to clarify transit timing variations of WASP-43~b, particularly regarding orbital decay.

Not only are the transit timing variations of WASP-43~b of interest, but extensive studies of its atmosphere have been conducted over the past decade. Focusing on WASP-43~b's emission spectra, the first eclipse data of WASP-43~b were obtained using the \textit{Spitzer Space Telescope's} Infrared Array Camera (IRAC) at 3.6 $\mu$m and 4.5 $\mu$m \citep{blecic2014}. The first complete orbital phase curve of WASP-43~b, obtained by \textit{Spitzer}, was presented by \citet{stevenson2017}, revealing the presence of H$ _{2}$O, CO+CO$_{2}$, and detecting variations in CH$_{4}$ emission spectra. More recently, \citet{lesjak2023} presented the first high-resolution dayside spectra of WASP-43~b and detected the presence of CO and H$_{2}$O abundances. Following the launch of the \emph{James Webb Space Telescope} (\emph{JWST}), \citet{bell2023} provided a full phase curve of WASP-43~b, observed using the Mid-Infrared Instrument Low Resolution Spectroscopy (MIRI/LRS) on the \emph{JWST}. The data were analyzed in \citet{bell2024}, showing a significant day–night temperature contrast in the spectrum and evidence of water absorption at all orbital phases.

For the transmission spectra of WASP-43~b, the first study of transmission spectra of WASP-43~b was conducted using the Wide Field Camera 3 (WFC3) on the \emph{Hubble Space Telescope} (\emph{HST}) \citep{kreidberg2014}, providing initial evidence for the presence of H$_{2}$O in its atmosphere. These \emph{HST} transmission data were later reanalyzed by \citet{tsiaras2018}, who identified evidence of H\(_{2}\)O, with a log volume mixing ratio of -4.36 \(\pm\) 2.10, and a cloud top pressure of 2.90 \(\pm\) 2.12. Moreover, \citet{weaver2020} presented ground-based transmission spectrum data for WASP-43~b from the ACCESS Survey; however, no significant presence of Na, K, or H$\alpha$ was detected. \citet{chubb2020} also re-analyzed the same \emph{HST} transmission data set, reporting evidence of AlO in transmission with a high significance level ($>5\sigma$) compared to a flat model. \citet{bartelt2025} observed four transits of WASP-43~b using the high-resolution Immersion GRating InfraRed Spectrometer (IGRINS) on the Gemini-S telescope, measuring a water abundance of $\log_{10}$(H$_2$O)=$-2.24^{+0.57}_{-0.48}$ without detecting any other carbon-bearing species.

In this work, we present a decade of ground-based multi-band photometric follow-up observations comprising 35 transits of WASP-43~b from the SPEARNET telescope network, a ground-based transmission spectroscopy survey \citep{hayes2024}. This extensive dataset enables robust constraints on the transit ephemeris and orbital decay. The details of these observations, combined with \emph{HST}, \emph{TESS}, \emph{JWST}, and other published data, are provided in Section~\ref{sec:observation}. Aligning with the objectives of SPEARNET, we perform a homogeneous analysis of all light curves using the \texttt{TransitFit} package to ensure parameter reliability and avoid inconsistencies arising from heterogeneous literature values, as detailed in Section~\ref{sec:LCModeling}. In Section~\ref{sec:ttv}, the derived mid-transit times are used to constrain a new linear ephemeris and investigate orbital decay and TTV signals. Additionally, the transit depths from the multi-band observations are analyzed to constrain the atmospheric composition of WASP-43~b in Section~\ref{sec:atmosphere}. Finally, the discussion and conclusion are presented in Section~\ref{sec:conclude}.

\section{Observational Data} 
\label{sec:observation}

In this work, we combined photometric light curve data spanning optical to mid-infrared wavelengths, including our observational data, published ground-based data, and data from \textit{HST}, \textit{TESS}, and \textit{JWST}. The details are provided in the following sections.

\subsection{SPEARNET Observations and Data Reduction}

Multi-band photometric follow-up observations of WASP-43~b were conducted between March 2017 and February 2022 using the SPEARNET telescope network \citep{hayes2024}. In total, transit photometry was obtained for 35 transits, comprising 28 full and 7 partial transits. The observation log and details of the SPEARNET facilities are provided in \Cref{tab:log}. Details of the specific telescopes used in this work are provided below:

\begin{enumerate}
    \item \emph{2.4-m Thai National Telescope (2.4-m TNT)} is located at the Thai National Observatory (TNO) in Thailand, we obtained multi-band photometric data for five full transits and one partial transit of WASP-43~b using the 2.4-m TNT between 2018 and 2021. The observations were performed with ULTRASPEC \citep{dhillon2014}, a high-speed frame-transfer EMCCD camera with 1024 $\times$ 1024 pixels, and a field-of-view of 7.68 $\times$ 7.68 arcmin$^{2}$. 

    \item \emph{0.5-m Thai Robotic Telescope at the Thai National Observatory (TRT-TNO), Thailand.} In 2017, we observed two full transits and one partial transit of WASP-43~b using the 0.5-m Schmidt-Cassegrain TRT-TNO. The observations were conducted with an Andor iKon-M 934 CCD camera (1024 $\times$ 1024 pixels), which features a field of view of approximately 23.4 $\times$ 23.4 arcmin$^{2}$.

    \item \emph{1-m Thai National Telescope (1-m TNT)} is located at the Thai National Observatory (TNO). The 1-m TNT is an upgraded version of the 0.5-m telescope at TNO. Starting in 2021, we observed six full transits of WASP-43~b using the 1-m Schmidt-Cassegrain telescope. The observations were conducted with an Andor iKon-M 934 CCD camera (1024 $\times$ 1024 pixels), which provides a field of view of approximately 23.4 $\times$ 23.4 arcmin$^{2}$.

    \item \emph{0.7-m Thai Robotic Telescope at the Gao Mei Gu Observatory (TRT-GAO), China.} Two full transit light curves of WASP-43~b were obtained by the TRT-GAO between 2017 and 2019. The observations were conducted using an Andor iKon-L 936 CCD camera equipped with a 2048 $\times$ 2048 pixels sensor. The camera provides a field of view of 20.9 $\times$ 20.9 arcmin$^{2}$.
    
    \item \emph{0.7-m Thai Robotic Telescope at the Spring Brook Obsrevatory (TRT-SBO), Australia.} WASP-43~b was observed by TRT-SBO between 2019 and 2021. During this period, six full transits and two partial transits were recorded. The TRT-SBO is equipped with a $4096 \times 4096$ pixel ProLine PL16803 Monochrome CCD camera, providing a field of view of 28 $\times$ 28 arcmin$^{2}$.
    
    \item \emph{0.7-m Thai Robotic Telescope at the Sierra Remote Observatories (TRT-SRO), USA.} One full transit light curve of WASP-43~b was obtained by TRT-SRO in 2022. The observation was conducted using an Andor iKon-M 934 CCD camera with a 1024 $\times$ 1024 pixel sensor, providing a field of view of 20.9 $\times$ 20.9 arcmin$^{2}$.
    
    \item \emph{0.7-m Regional Observatory for the Public Nakhon Ratchasima (ROP-NM), Thailand.} During 2017-2018, the 0.7-m ROP-NM obtained four full and three partial transit light curves of WASP-43~b. The observations were conducted using a $4096 \times 4096$ pixel ProLine PL16803 Monochrome CCD camera, with a field of view of approximately 28 $\times$ 28 arcmin$^{2}$.
    
    \item \emph{0.7-m Regional Observatory for the Public Chachoengsao (ROP-CC), Thailand.} Two full transit light curves were obtained by the 0.7-m ROP-CC from 2018 to 2019. The observations were conducted using the ProLine PL16803 Monochrome CCD camera ($4096 \times 4096$ pixels), providing a field of view of approximately 28 $\times$ 28 arcmin$^{2}$.
\end{enumerate} 

The SPEARNET multi-band photometric observations were calibrated using standard tasks from {\tt IRAF}\footnote{IRAF is distributed by the National Optical Astronomy Observatories, which are operated by the Association of Universities for Research in Astronomy, Inc., under a cooperative agreement with the National Science Foundation (\texttt{http://iraf.noao.edu/}).} \citep{tody1986,tody1993}. Astrometric calibrations were performed using {\tt Astrometry.net} \citep{lang2010}. Aperture photometry was carried out on all images using {\tt Source Extractor} \citep{bertin1996}. Reference stars were chosen from nearby stars within 3 magnitudes of WASP-43 that exhibited no variations in brightness. Outlier points in the light curves were removed using a 5-sigma clipping algorithm. To generate the light curves, the flux of WASP-43 was divided by the sum of the flux from the selected reference stars. The time stamps were converted to Barycentric Julian Date in Barycentric Dynamical Time (BJD$_\textup{TDB}$) using {\tt barycorrpy} \citep{Kano2018}. The normalized light curves are available in a machine-readable format in \Cref{tab:lightcurve}.

\begin{table*}
\begin{center}
\caption{Observation log of WASP-43~b transits observed with the SPEARNET telescopes.}
\label{tab:log}          
\small\addtolength{\tabcolsep}{-2pt}
\begin{tabular}{lcccccccc}
\toprule
\multirow{2}{*}{Observation date}  & \multirow{2}{*}{Epoch$^*$} & \multirow{2}{*}{Telescope}  & \multirow{2}{*}{Filter} & \multirow{2}{*}{Exposure time (s)} & Number & Total duration of & \multirow{2}{*}{PNR (\%)$^\dagger$ } & Transit \\
  &   &   &   &  & of images & observation (hr) &   &  coverage \\
\hline
2017 Mar 04	&	484	&	0.7-m ROP-NM	&	$R$	&	20	&	128	&	2.40	&	0.25	&	Full	\\
	&	484	&	0.7-m ROP-NM	&	$V$	&	35	&	131	&	2.45	&	0.37	&	Full	\\
2017 Mar 13	&	495	&	0.7-m ROP-NM	&	$R$	&	60	&	111	&	2.59	&	0.34	&	Full	\\
	&	495	&	0.7-m TRT-GAO	&	$R$	&	20	&	102	&	0.70	&	0.25	&	Full	\\
	&	495	&	0.5-m TRT-TNT	&	$I$	&	20	&	220	&	1.72	&	0.49	&	Egress only	\\
2017 Mar 21	&	505	&	0.7-m ROP-NM	&	$R$	&	20	&	270	&	2.21	&	0.18	&	Full	\\
	&	505	&	0.5-m TRT-TNT	&	$I$	&	20	&	258	&	1.76	&	0.41	&	Full	\\
2017 Mar 30	&	516	&	0.5-m TRT-TNT	&	$I$	&	20	&	461	&	2.89	&	0.35	&	Full	\\
2018 Jan 29	&	891	&	2.4-m TNT	&	$g'$	&	12.83	&	864	&	3.08	&	0.14	&	Full	\\
2018 Feb 02	&	896	&	2.4-m TNT	&	$r'$	&	3.56	&	535	&	1.60	&	0.17	&	Ingress only	\\
2018 Feb 07	&	902	&	0.7-m ROP-NM	&	$V$	&	70	&	17	&	1.99	&	0.25	&	Egress only	\\
	&	902	&	0.7-m ROP-NM	&	$R$	&	70	&	17	&	1.35	&	0.28	&	Egress only	\\
	&	902	&	0.7-m ROP-NM	&	$I$	&	70	&	23	&	1.86	&	0.32	&	Egress only	\\
2018 Feb 16	&	913	&	0.7-m ROP-CC	&	$R$	&	30	&	217	&	2.13	&	0.39	&	Full	\\
2018 Mar 01	&	926	&	2.4-m TNT	&	$i'$	&	2.16	&	3289	&	2.81	&	0.17	&	Full	\\
2018 Mar 23	&	956	&	2.4-m TNT	&	$z'$	&	12.78	&	850	&	3.12	&	0.15	&	Full	\\
2019 Jan 21	&	1330	&	2.4-m TNT	&	$r'$	&	3.51	&	2482	&	2.43	&	0.09	&	Full	\\
2019 Feb 09	&	1353	&	0.7-m TRT-SBO	&	$I$	&	60	&	232	&	5.46	&	0.32	&	Full	\\
2019 Feb 13	&	1358	&	0.7-m TRT-SBO	&	$I$	&	60	&	78	&	1.73	&	0.31	&	Full	\\
2019 Feb 22	&	1369	&	0.7-m ROP-CC	&	$V$	&	30	&	315	&	3.74	&	0.37	&	Full	\\
2019 Mar 07	&	1385	&	0.7-m TRT-SBO	&	$I$	&	60	&	73	&	3.62	&	0.22	&	Full	\\
2019 Apr 11	&	1428	&	0.7-m TRT-GAO	&	$R$	&	60	&	170	&	3.97	&	0.22	&	Full	\\
2019 Apr 16	&	1434	&	0.7-m TRT-SBO	&	$I$	&	60	&	183	&	3.77	&	0.30	&	Full	\\
2020 Dec 11	&	2178	&	0.7-m TRT-SBO	&	$R$	&	40	&	126	&	1.93	&	0.25	&	Full	\\
2021 Jan 18	&	2225	&	1-m TNT	&	$R$	&	3	&	707	&	2.06	&	0.15	&	Full	\\
2021 Jan 19	&	2226	&	0.7-m TRT-SBO	&	$R$	&	40	&	125	&	1.99	&	0.21	&	Ingress only	\\
2021 Jan 23	&	2231	&	1-m TNT	&	$R$	&	3	&	857	&	2.67	&	0.12	&	Full	\\
2021 Feb 09	&	2252	&	1-m TNT	&	$R$	&	3	&	1031	&	3.12	&	0.14	&	Full	\\
2021 Feb 10	&	2253	&	0.7-m TRT-SBO	&	$R$	&	31	&	73	&	3.11	&	0.21	&	Full	\\
2021 Feb 15	&	2259	&	0.7-m TRT-SBO	&	$R$	&	31	&	26	&	1.56	&	0.12	&	Ingress only	\\
2021 Mar 17	&	2296	&	1-m TNT	&	$R$	&	3	&	652	&	1.83	&	0.12	&	Full	\\
	&	2296	&	2.4-m TNT	&	$r'$	&	2.19	&	2774	&	1.70	&	0.20	&	Full	\\
2021 Dec 07	&	2622	&	1-m TNT	&	$V$	&	5	&	1201	&	3.60	&	0.14	&	Full	\\
2021 Dec 20	&	2638	&	1-m TNT	&	$V$	&	5	&	943	&	3.00	&	0.15	&	Full	\\
2022 Feb 09	&	2700	&	0.7-m TRT-SRO	&	$R$	&	30	&	208	&	2.53	&	0.37	&	Full	\\
\hline
\end{tabular}
\end{center}
{\textbf{Note}: $^*$ Epoch=0 is the transit on 2016 February 4. $^\dagger$ PNR is the photometric noise rate \citep{Fulton2011}.}
\end{table*}

\subsection{Existing Ground-based Data}

After its discovery in 2011, WASP-43~b has been observed by a number of ground-based telescopes. In addition to the light curves from our observations, we utilized 109 ground-based transit light curves from eight previous studies. The published light curves, along with details of their instruments and filters, are listed in \Cref{tab:log_achived}.

\begin{table*}
\begin{center}
\caption{WASP-43~b's transit light curves taken from available published data for the analyses.}
\label{tab:log_achived}          
\small\addtolength{\tabcolsep}{-2pt}
\begin{tabular}{lcccc}
\toprule
\multirow{2}{*}{ Sources} & \multirow{2}{*}{Telescope}  & \multirow{2}{*}{Filter} & Number of  & Epoch \\
& & & light curve  & span$^*$ \\
\hline
 \multirow{2}{*}{\citet{gillon2012}} & TRAPPIST & $I+z'$& 19 & \multirow{2}{*}{(-2086, -2318)}\\
              & Euler   & Gunn-$r'$ & 3 \\
\hline
 \multirow{2}{*}{\citet{maciejewski2013}} & 2.2-m Calar Alto & $R$& 1 & -1789 \\
              & 0.6-m Toru$\grave{n}$  & $Clear$& 1  & -1297 \\
\hline
\multirow{7}{*}{\citet{chen2014}} & \multirow{7}{*}{{MPG/ESO 2.2-m telescope}}  &$g'$& 1 & \multirow{7}{*}{-1830}\\
 &  &$r'$& 1 \\
 &  &$i'$& 1 \\
 &  &$z'$& 1 \\
 &  &$J$& 1 \\
 &  &$H$& 1 \\
 &  &$K$& 1 \\
\hline
\citet{murgas2014} & 10.4-m GTC/OSIRIS & 0.52 - 1.04 $\mu$m & 34  & -1379 \\
\hline
\multirow{4}{*}{\citet{ricci2015}} & 1.5-m San Pedro M$\grave{a}$rtir telescopes &$i'$& 2  & \multirow{4}{*}{(-779,-887)}\\
& \multirow{3}{*}{{0.84-m San Pedro M$\grave{a}$rtir telescopes}} &$V$& 2 \\
&  &$R$& 2 \\
&  &$I$& 2 \\
\hline
\citet{jiang2016} & P60 Palomar Observatory& $R$ & 7  & (-469,-854)\\
\hline
\multirow{4}{*}{\citet{parvia2019}} & \multirow{4}{*}{{MuSCAT2}}  &$g'$& 3 & \multirow{4}{*}{(867,970)}\\
&   &$r'$& 3 \\
&   &$i'$& 2 \\
&   &$z'$& 3 \\
\hline
\multirow{4}{*}{\citet{garai2021}} & \multirow{4}{*}{{MuSCAT2}}   &$g'$& 4 & \multirow{4}{*}{(916,1307)}\\
&   &$r'$& 5 \\
&  &$i'$& 5 \\
&   &$z'$& 4 \\
\hline
\multicolumn{2}{c}{\emph{TESS} Sector No. 09} & \emph{TESS} & 26  & (1379, 1407)\\
\multicolumn{2}{c}{\emph{TESS} Sector No. 35} & \emph{TESS} & 24 & (2253, 2282) \\
\multicolumn{2}{c}{\emph{TESS} Sector No. 62} & \emph{TESS} & 28 & (3155, 3184) \\
\multicolumn{2}{c}{\emph{TESS} Sector No. 89} & \emph{TESS} & 33 & (4051, 4083) \\

\hline
\multicolumn{2}{c}{\multirow{2}{*}{\emph{HST}/WFC3}} & G141 & \multirow{2}{*}{150} & \multirow{2}{*}{(-993, -1011)} \\
& & (1.1 - 1.7 $\mu$m.)  & \\
\hline
\multicolumn{2}{c}{\multirow{2}{*}{\emph{JWST}/MIRI}} & P750L & \multirow{2}{*}{11} & \multirow{2}{*}{3063} \\
& & (5 - 12 $\mu$m.)  & \\
\hline
\end{tabular}
\end{center}
{\textbf{Note}: $^*$ Epoch=0 is the transit on 2016 February 4.}
\end{table*}

\subsection{\emph{HST}/WFC3 G141 Grism Data}
WASP-43~b was observed by \emph{HST}/WFC3 as part of Proposal ID 13467 (P.I. Jacob Bean) \citep{kreidberg2014}. During the observation period from November 4 to 19, 2013, six transits were observed using the G141 grism, covering the wavelength range from 1.1 to 1.7 $\mu$m. In this work, the \emph{HST}/WFC3 raw spectra of WASP-43~b were downloaded from the Exo.MAST\footnote{Downloaded from Exo.MAST: \texttt{https://exo.mast.stsci.edu/}} database. The data reduction was performed using the \texttt{Iraclis} package, a Python tool for the WFC3 spectroscopic reduction pipeline \citep{tsiaras-waldmann2016, tsiaras2016}. The G141 grism spectra were binned into 25 wavelength bins, resulting in a total of 150 light curves obtained from \emph{HST}/WFC3. Data from the first orbit of each visit and the first exposure of each orbit were discarded due to the presence of a stronger wavelength-dependent ramp during these epochs \citep{tsiaras2016}. 
 
\subsection{\emph{TESS} Data}
\emph{TESS} observed WASP-43~b in four time intervals, yielding a total of 26 transit light curves from Sector No. 09 (2019 February 28 - March 26), 24 light curves from Sector No. 35 (2021 February 9 - March 7), 28 light curves from Sector No. 62 (2023 February 12 - March 9) and 33 light curves from Sector No. 89 (2025 February 12 - March 10). These light curves were downloaded from the Mikulski Archive for Space Telescopes (MAST)\footnote{Downloaded from the Mikulski Archive for Space Telescopes: \texttt{https://archive.stsci.edu/}}. We used the Pre-Search Data Conditioning (PDC) light curves, which are the calibrated data provided by the Science Processing Operation Center (SPOC) pipeline \citep{jenkins2016}. 

\subsection{\emph{JWST} Data}
A full orbit of WASP-43~b was observed by \emph{JWST} on 2022 December 1 as part of the Transiting Exoplanet Community Early Release Science Program (JWST-ERS-1366), led by PI: Taylor J. Bell \citep{bell2023, bell2024}. The observation utilized the \emph{JWST} Mid-Infrared Instrument (MIRI) in Low-Resolution Spectroscopy (LRS) slitless mode with the P750L filter, covering a wavelength range from 5 to 12 $\mu$m. We used the 11 available \emph{JWST} light curves reduced by \texttt{Eureka!v1} \citep{bell2022} from \citet{bell2024}.

\section{TransitFit Light-Curve Modeling}
\label{sec:LCModeling}

To determine the planetary parameters of WASP-43~b, we used \texttt{TransitFit}, a Python package designed for fitting multi-filter and multi-epoch exoplanet transit observations simultaneously \citep{hayes2024}. The package employs the transit model from \texttt{batman} \citep{kreidberg2015} and utilizes dynamic nested sampling routines from \texttt{dynesty} \citep{speagle2020} to derive the parameters.

As mentioned in Section~\ref{sec:observation}, we utilized a large number of light curves from both ground-based and space-based observations. To mitigate issues related to computer memory, we divided the light curve data into four distinct groups: Ground-based, \emph{HST}, \emph{TESS}, and \emph{JWST} datasets. The light curves from each group were simultaneously fitted and detrended using \texttt{TransitFit}.

For the light curve detrending, each transit light curve was individually detrended using different detrending functions. We applied second-order polynomial detrending functions for the ground-based, \emph{TESS}, and \emph{JWST} datasets. The normalized light curves from our SPEARNET ground-based observations are shown in \Cref{tab:lightcurve}. For the \emph{HST}/WFC3 G141 data, we used a custom detrending function implemented in \texttt{TransitFit}. This detrending function was based on the method described by \citet{kreidberg2018}, specifically designed for the WFC3 data,
\begin{equation}
    F_{\textup{sys}} = (S + v_{1}t_{\textup{visit}} + v_{2}t^{2}_{\textup{visit}})(1 - e^{-at_{\textup{orb}}-b}) \ ,
\end{equation}
where $F_\textup{sys}$ is the signal from the systematics, where $S = 1$ and $S = s$ for forward and reverse scans, respectively. The parameters $s$, $v_{1}$, $v_{2}$, $a$, and $b$ are detrending coefficients, with $s$, $a$, and $b$ accounting for the ramp-up systematic across all the light curves, while $v_{1}$ and $v_{2}$ are the second-order polynomial detrending functions used to model the visit-long trends. The time elapsed since the first exposure in the visit is represented as $t_{\textup{visit}}$, and the time elapsed since the first exposure in the orbit is presented as $t_{\textup{orb}}$. The astrophysical signal ($F_\textup{\textup{sig}}$) can be obtained by dividing the observed flux ($F_\textup{\textup{obs}}$) by the systematic signal ($F_\textup{\textup{sys}}$).

To perform the \texttt{TransitFit} fitting, we used a stellar effective temperature for the host star of WASP-43~b of $T=4166\pm100$ K, calculated using the Python Stellar Spectral Energy Distribution package\footnote{Python Stellar Spectral Energy Distribution package on \url{https://explore-platform.eu/}}. This toolset is designed to allow the user to create, modify, and fit the spectral energy distributions of stars based on publicly available data \citep{mcdonald2009,mcdonald2012,mcdonald2017}. The host metallicity, $Z_{*} = -0.05\pm0.17$, from \citet{bonomo2017}, and surface gravity, $\textup{log} (g_{*}) = 4.6\pm0.1$, from the Gaia EDR3 catalogue\footnote{Gaia archive: \texttt{https://archives.esac.esa.int/gaia}}, were used. To find the best fit for all light curves from the four different data sets, we assumed that the orbit of WASP-43~b is circular. The priors for each parameter: orbital period ($P$ in days), mid-transit for each epoch (${T}_{m}$ in BJD), orbital inclination ($i$ in deg), semimajor axis ($a$ in units of stellar radius, $R_{*}$), and the planet's radius ($R_p$ in units of stellar radius, $R_{*}$), for each filter are given in \Cref{tab:initialpara}.

Since WASP-43~b was observed over several years, its orbital period might not be constant. Therefore, we first determined the best value for the orbital period $P$. The prior for the orbital period, with a Gaussian distribution of $0.813474 \pm 1 \times 10^{-6}$ day, was used to calculate the best orbital period for each dataset. The parameters for inclination $i$ and semimajor axis $a$ were allowed to vary. The best-fit values for the inclination, semimajor axis, and orbital period, obtained from the analysis of each dataset, are given in \Cref{tab:outpara}. From the analysis, the best-fit values from each dataset are combined using the weighted mean. The weighted mean parameters of WASP-43~b indicate an inclination of $82.12 \pm 0.01$ degrees and a semimajor axis of $4.845 \pm 0.003R_{*}$. A comparison shows that the results of our study are compatible with previous measurements within $1\sigma$.

The investigation of Transit Timing Variations (TTVs) was conducted using the \texttt{allow\_TTV} function in \texttt{TransitFit}. In this step, the weighted mean values of the orbital period, inclination, and semimajor axis, determined in the first step, were fixed to account for TTVs and the mid-transit time ($T_m$) for each epoch. The planet's radius ($R_p$) and limb-darkening coefficients (LDCs) were allowed to vary. For the analysis of the LDCs for each filter, the fitting was performed using the \texttt{custom} LDCs fitting mode in \texttt{TransitFit}. The prior LDC values for each filter were obtained from the Exoplanet Characterization Toolkit (ExoCTK, \citet{ExoCTK})\footnote{ExoCTK limb darkening calculator: \texttt{https://exoctk.stsci.edu/limb\_darkening}}. However, for the \emph{HST} light curve fitting, the LDCs were fixed due to the gap between the orbital ingress and egress parts of the light curve. The limb-darkening parameters calculated using \texttt{TransitFit} show the same trend as those obtained from \texttt{ExoCTK}, with a difference of less than 0.2 for the first-order limb-darkening coefficient ($u_0$).

\begin{figure*}[htb]
\centering
    \includegraphics[width=0.9\textwidth]{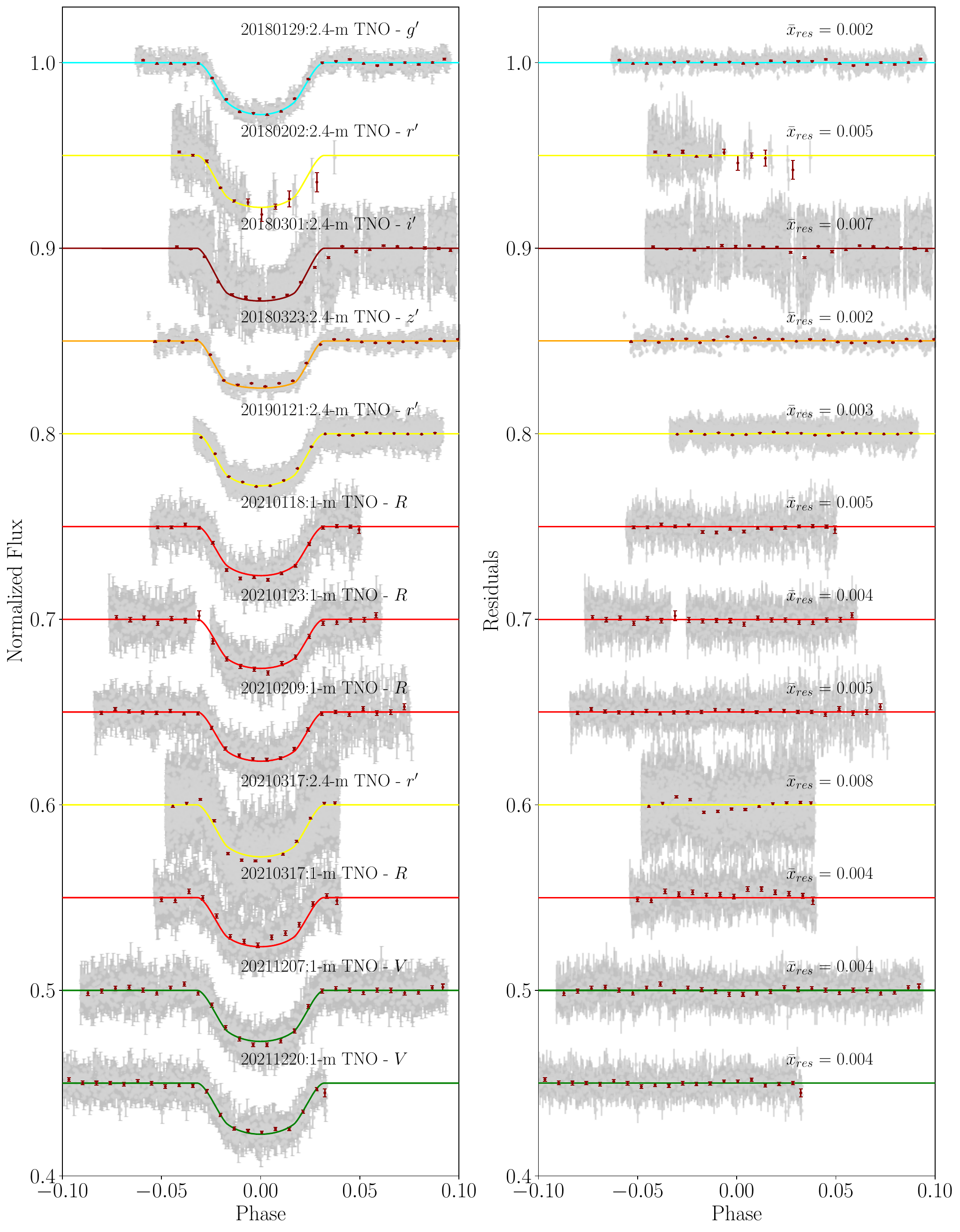}
    \caption{{\it Left panels:} The normalized, phase-folded transit light curves of WASP-43~b from the 2.4-m and 1-m TNT observations in the SPEARNET telescope network, shown as grey dots. The best-fitting model from \texttt{TransitFit} is displayed as a solid line. {\it Right panels:} Residuals of the light curves after model subtraction. Both the light curves and residuals are vertically offset for clarity. The corresponding residuals and the mean residual values (${\bar{x}_{res}}$) with clear offsets are displayed on the right panel.}
\label{fig:LCs_TNT}
\end{figure*}

\begin{figure*}[htb]
\centering
  \begin{tabular}{cc}
    \includegraphics[width=0.49\textwidth,height=0.572\textwidth,page=1]{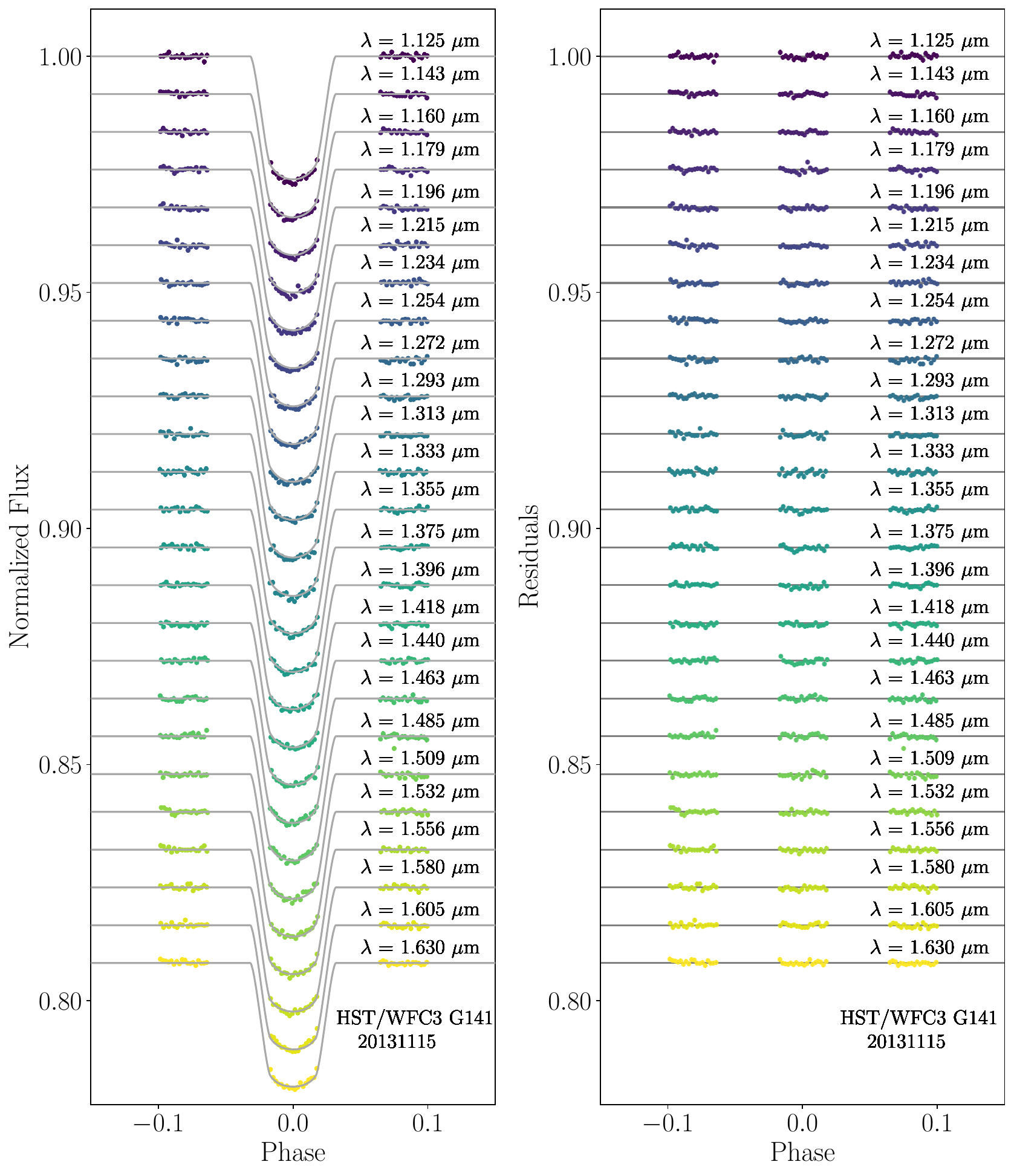} 
    \includegraphics[width=0.49\textwidth,page=1]{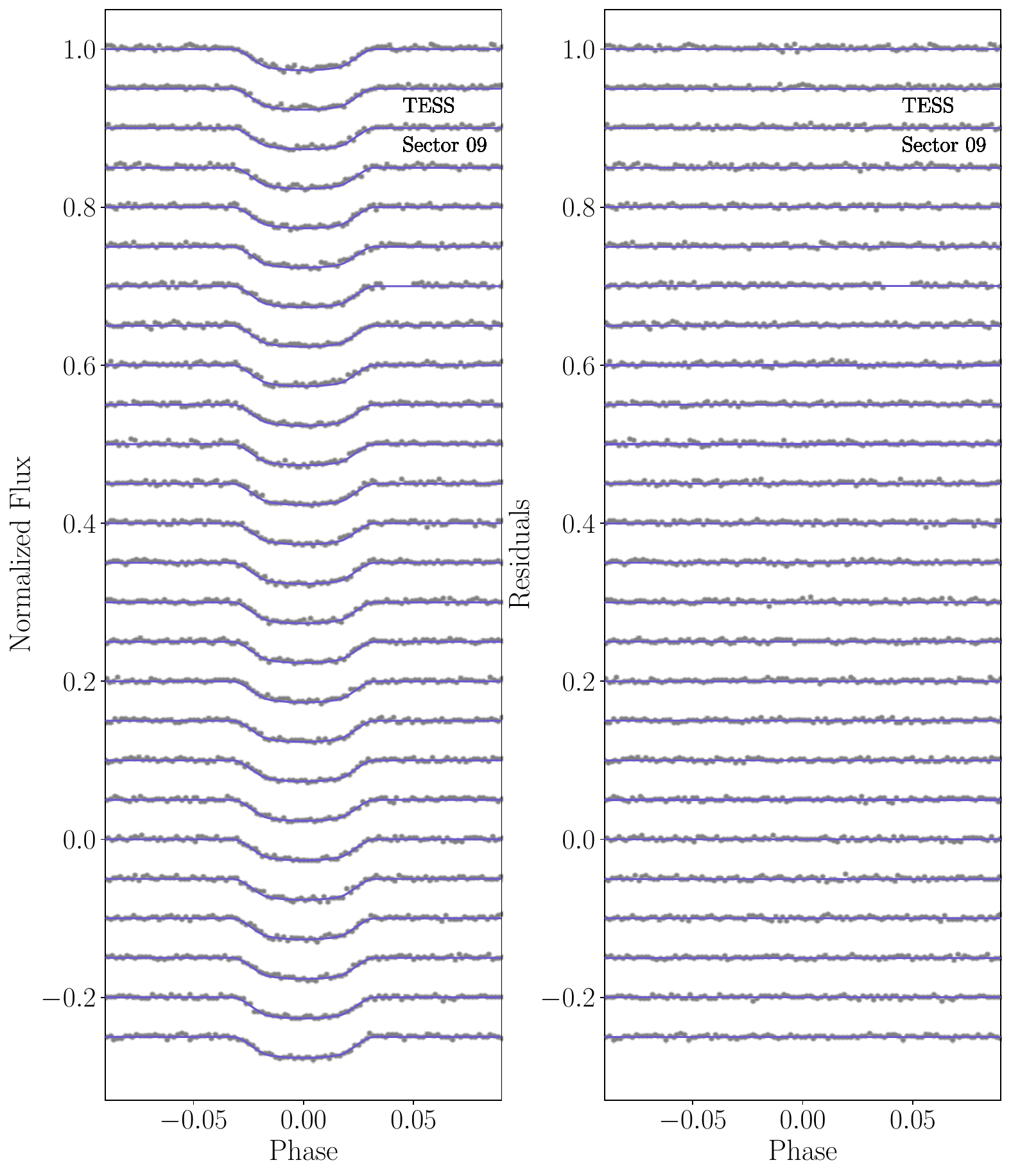} \\
    \includegraphics[width=0.49\textwidth,page=1]{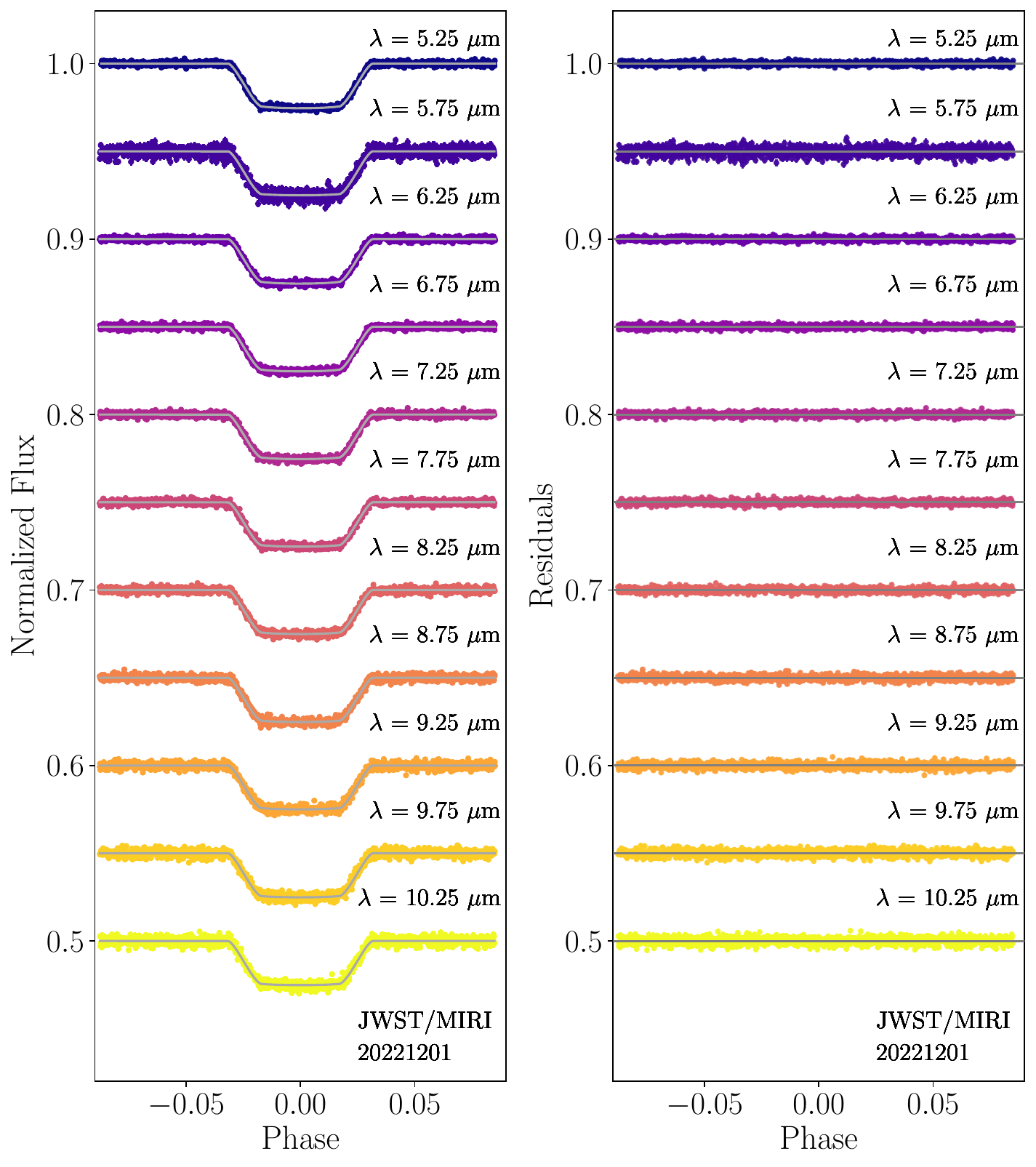} 
    \end{tabular}
    \caption{The normalized, phase-folded transit light curves of WASP-43~b are shown for \emph{HST}/WFC3 G141 observations on November 15, 2013 (upper left), \emph{TESS} observations within Sector 9 (upper right), and \emph{JWST}/MIRI (bottom). The observational data are presented as dots, and the best-fitting model from \texttt{TransitFit} is shown as solid lines. Both the light curves and the corresponding residuals (right panels) are vertically offset for clarity.}
\label{fig:LCs_HTJ}
\end{figure*}

The light curves were phase-folded with their best-fit models, and the residuals are shown in \Cref{fig:LCs_TNT,fig:LCs_HTJ}. The individual light curve fittings for 23 TRT light curves, \emph{HST} observations from November 04–19, 2013, and \emph{TESS} observations in Sectors 35, 62 and 89 are presented in Appendix D. The mid-transit times are provided in \Cref{tab:midtransit} and discussed in Section~\ref{sec:ttv}. The LDCs were fitted as free parameters using the quadratic limb-darkening model. The values of the limb-darkening coefficients for 58 different filters, derived from \texttt{TransitFit}, are listed in \Cref{tab:radius-transitDepth-limbdark}.

\begin{table*}
\begin{center}
\caption{The detrended transit light curves of WASP-43~b, observed by the telescopes within the SPEARNET network.}
\label{tab:lightcurve}          
\small\addtolength{\tabcolsep}{-2pt}
\begin{tabular}{lcccc}
\toprule
\multirow{2}{*}{Epoch} & \multirow{2}{*}{BJD} & \multirow{2}{*}{Normalized Flux} & Normalized flux   \\
      &     &                 & Error       \\
\hline
484	&	2457817.13376	&	1.001	&	0.007	\\
    &	2457817.13442	&	0.996	&	0.007	\\
    &	2457817.13508	&	1.002	&	0.006	\\ 
   & ...   & ...    & ...        \\
\hline							
495	&	2457826.08193	&	0.988	&	0.005	\\
    &	2457826.08220	&	0.999	&	0.005	\\
    &	2457826.08248	&	0.994	&	0.005	\\
    & ...   & ...    & ...        \\
\hline							
505	&	2457834.22363	&	1.000	&	0.003	\\
    &	2457834.22440	&	0.993	&	0.002	\\
    &	2457834.22466	&	0.998	&	0.002	\\
    & ...   & ...    & ...        \\
\hline
... & ...   & ...    & ...        \\
\hline
\end{tabular}
\end{center}
{\textbf{Note}: The full table is available in a machine-readable format in the online journal.}
\end{table*}

\begin{table*}
\begin{center}
\caption{The priors used to model the planetary parameters of WASP-43~b for the analysis of all ground-based, \emph{TESS}, \emph{JWST}, and \emph{HST}/WFC3 light curves with \texttt{TransitFit}.}
\label{tab:initialpara}          
\small\addtolength{\tabcolsep}{-2pt}
\begin{tabular}{lcc}
\toprule
Parameter  &  Priors &  Prior distribution \\
\hline
$P$ (days)      &  $0.813474 \pm 1\times10^{-6}$   &  Gaussian   \\
$t_{0}$ (BJD) &  2457423.45 $\pm$ 0.002	       &  Gaussian  \\
$i$ (deg)       &  (80, 84)                        &  Uniform  \\
{$a/R_{*}$}     &  (4,6)   &   Uniform  \\
{$R_p$/$R_{*}$} &    (0.14, 0.17)   &   Uniform     \\
$e$             & 0     & Fixed     \\
$T_{*}$ (K)     &     $4166 \pm 100$   & Fixed     \\
$Z_{*}$         &     $-0.05 \pm 0.17$  & Fixed     \\
$\textup{log}~(g_*)$  &  $4.6 \pm 0.1$  & Fixed     \\
\hline
\end{tabular}
\end{center}
{\textbf{Note.} The priors of $P$, $i$, and {$a/R_{*}$} are set to the values in \citet{hellier2011}, and ${t}_{0}$ is set to the value in \citet{ivshina2022}.}
\end{table*}

\begin{table*}
\begin{center}
\caption{The best-fit values of the planetary parameters of WASP-43~b for all four datasets modeled with \texttt{TransitFit}.}
\label{tab:outpara}          
\small\addtolength{\tabcolsep}{-2pt}
\begin{tabular}{cccc}
\toprule
Reference   & $P$ (days) & $i$ (deg)  &  {$a/R_{*}$}\\
\hline
\citet{hellier2011} & 0.813475 $\pm$ 1 $\times$ 10$^{-6}$ & $82.6^{+1.3}_{-0.9}$ & 4.97  $\pm$ 0.14 \\
\citet{bonomo2017} & 0.81347437 $\pm$ 1.3 $\times$ 10$^{-7}$ & 82.33 $\pm$ 0.20  & - \\
\citet{kokori2023} & 0.813474056 $\pm$ 2.1 $\times$ 10$^{-8}$ & 82.11 $\pm$ 0.10  & 4.867 $\pm$ 0.023 \\
\hline
\multicolumn{4}{c}{This study}   \\
\hline
Ground-Based &  0.81347420 $\pm$ 1$\times$10$^{-8}$ & 81.89	$\pm$ 0.02 & 4.799 $\pm$ 0.005  \\
\emph{HST} & 0.81347442 $\pm$ 7$\times$10$^{-8}$ &  82.35 $\pm$ 0.02 & 4.896 $\pm$ 0.005   \\
\emph{TESS} & 0.81347406 $\pm$ 1$\times$10$^{-8}$ & 82.03 $\pm$	0.06 & 4.83 $\pm$ 0.02   \\
\emph{JWST} & 0.8134739 $\pm$ 1$\times$10$^{-7}$ & 82.04 $\pm$ 0.02 & 4.839 $\pm$ 0.004  \\
Weighted  & \multirow{2}{*}{0.813474131 $\pm$ 7$\times$10$^{-9}$} & \multirow{2}{*}{82.12 $\pm$ 0.01} &  \multirow{2}{*}{4.845 $\pm$ 0.003}  \\
Mean Values &  &  &    \\
\hline
\end{tabular}
\end{center}
\end{table*}

\section{Timing Analysis}
\label{sec:ttv}

\subsection{Ephemeris Refinement}
To search for timing variations in WASP-43~b, the mid-transit times of 188 epochs obtained from \texttt{TransitFit}, listed in \Cref{tab:midtransit}, were analyzed. First, a new linear ephemeris was derived by fitting the mid-transit times with a constant-period model as follows: 
\begin{equation}
t_{0,c} =  t_{0,l} + P_{l} \times E \ ,
\end{equation}
where $t_{0,l}$ and $P_{l}$ are the reference time and the orbital period of the linear ephemeris model, respectively, $E$ is the epoch number and $E = 0$ represents the transit on 2016 February 4. $t_{0,c}$ is the calculated mid-transit time at a given epoch $E$.

The fitting was conducted using \texttt{emcee} \citep{foreman2013}, which employs a Markov Chain Monte Carlo (MCMC) algorithm with 50 chains and $10^{5}$ MCMC steps to determine the optimal parameters for the model. A summary of the priors used and the best-fit model parameters is presented in \Cref{tab:timing}. The new defined linear ephemeris is as follows:
\begin{equation}  
\label{eq:linear1}
t_{0,c} =  2457423.44976^{+0.00003}_{-0.00003} + 0.81347406^{+1\times10^{-8}}_{-1\times10^{-8}} E\ .
\end{equation}

The reduced chi-square for the linear model is $\chi^{2}_{\rm red} = 18.8$ with 186 degrees of freedom. The Bayesian Information Criterion (BIC) is calculated as $BIC = \chi^{2} + k \ln n = 3511$, where $k$ represents the number of free parameters, and $n$ is the number of data points. The corner plot of the MCMC posterior distribution is shown in \Cref{fig:liDe_mcmc}. Using this new linear ephemeris, we constructed the $O-C$ diagram of WASP-43~b (\Cref{fig:oc}), which displays the timing residuals ($O-C$) between the observed timing data and the linear equation (Equation 3.).

\subsection{Orbital Decay Investigation}
WASP-43~b remains a good candidate for observing orbital decay due to its ultra-short orbital period, although \citet{garai2021} have highlighted the orbital decay of WASP-43~b with unresolved conclusions. The orbital decay of WASP-43~b was investigated in this work. The timing data for a total of 188 epochs were also fitted with the orbital decay model using the following equation:
\begin{equation}
t_{0,c} = t_{0,d} + E \times P_{d} + \frac{1}{2}\frac{\textup{d} P_{d}}{\textup{d} E} E^2 \ ,
\end{equation}
where $t_{0,d}$ is a reference time of the orbital decay model. $P_{d}$ is planetary orbital period of the orbital decay model and $\textup{d}P_{d}/\textup{d}E$ is the change of orbital in each orbit. 

\begin{figure*}[htb]
\centering
    \includegraphics[width=0.6\textwidth,page=1]{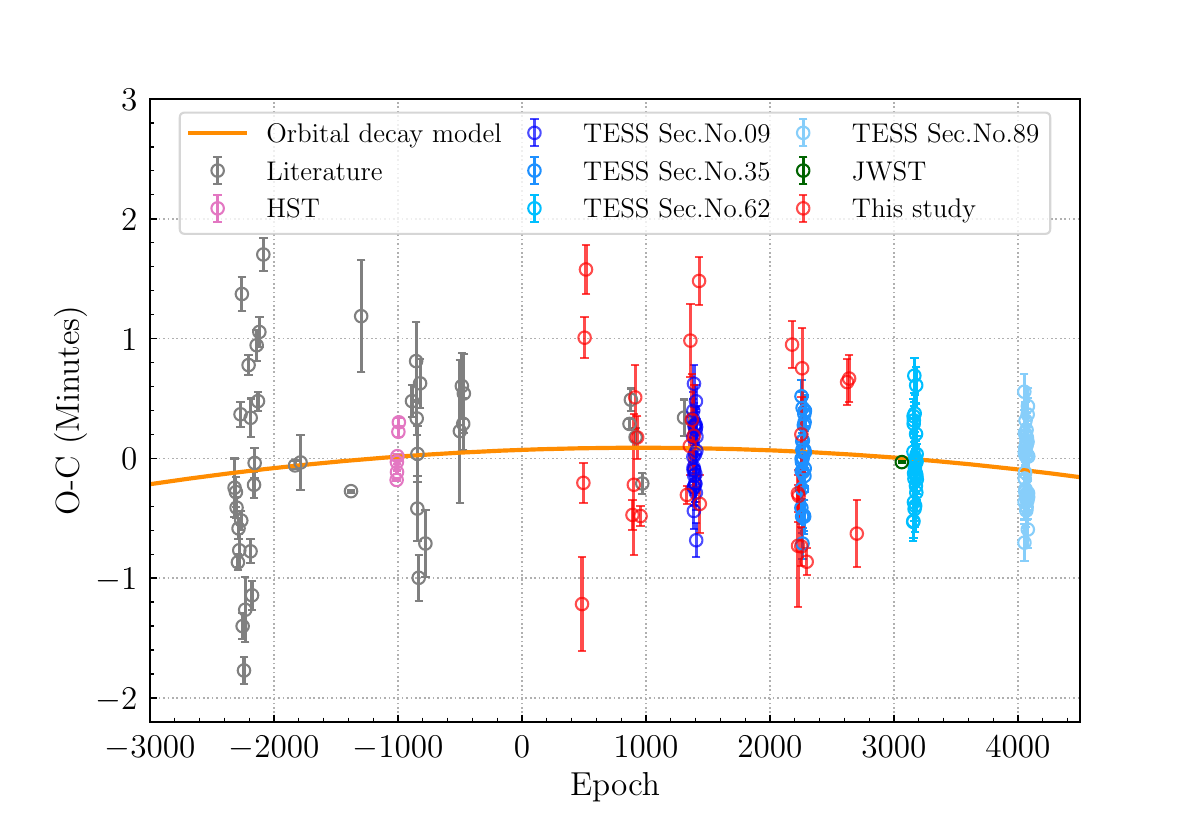} 
    \caption{The $O-C$ diagram and best-fitting timing models for WASP-43~b, including data from the literature (grey circles), \emph{HST} (pink circles), \emph{TESS} (blue circles), \emph{JWST} (green circles), and this study (red circles), are shown. The orange line represents the timing residuals of the orbital decay model.}
    \label{fig:oc}
\end{figure*}

The fitting for the orbital decay model was performed using the MCMC routine. The priors used and the best-fitting model are provided in \Cref{tab:timing}, with the posterior distribution of the MCMC shown in \Cref{fig:liDe_mcmc}. From the best-fit parameters, the timing residuals as a function of epoch $E$ for the orbital decay model, obtained by subtracting the best-fitting constant-period model, are shown in \Cref{fig:oc}. We obtained the change in the orbital period, $\textup{d}P_{d}/\textup{d}E = -3^{+1}_{-1} \times 10^{-11}$ days/orbit, with the reduced chi-square of the model ${\chi}^{2}_{\textup{red}} = 18.3$ (185 degrees of freedom) and BIC = 3408.

Since the values of ${\chi}^{2}_{\textup{red}}$ and BIC from both the constant-period model and the orbital decay model in our analysis do not show a significant difference, there is no strong evidence for the detection of orbital decay in WASP-43~b.

\begin{table*}
\begin{center}
\caption {The priors used, the uniform distribution, and the best-fitting parameters from the MCMC for both the constant-period and orbital decay models.}
\label{tab:timing}
\begin{tabular}{lcc}
\hline
\hline
Parameter & Uniform distribution priors & Best fit values  \\ 
\hline
\multicolumn{3}{c}{Constant-period model} \\
\hline
$P_{\textup{orb},l}$ [days]          &  (0.81347,0.81348)    & $0.81347406
^{+1\times10^{-8}}_{-1\times10^{-8}}$ \\
$t_{0,l}$ [BJD$_{\textup{TDB}}$]     &  (2457423.445, 2457423.453)   &   $2457423.44976
^{+0.00003}_{-0.00003}$  \\
${\chi}^{2}_{\textup{red}}$ &  \multicolumn{2}{c}{18.8} \\
BIC & \multicolumn{2}{c}{3511} \\
\hline
\multicolumn{3}{c}{Orbital decay model} \\
\hline
$P_{\textup{orb},d}$ [days]                &  (0.81347, 0.81348)    &   $0.81347409
^{+2\times10^{-8}}_{-2\times10^{-8}}$   \\
$t_{0,d}$ [$BJD_{\textup{TDB}}$]         &  (2457423.445, 2457423.453)   &  $2457423.44981
 ^{+0.00004}_{-0.00004}$  \\
$dP/dE$ [days/orbit]            &  (-0.2, 0.2) &   $-3^{+1}_{-1}\times10^{-11}$  \\
${\chi}^{2}_{\textup{red}}$ &  \multicolumn{2}{c}{18.3} \\
BIC & \multicolumn{2}{c}{3408} \\
\hline
\end{tabular}\\
\end{center}           
\end{table*}

\subsection{Searching for the Periodical in Timing Data}

\begin{figure*}[htb]
\centering
    \includegraphics[width=0.5\textwidth,page=1]{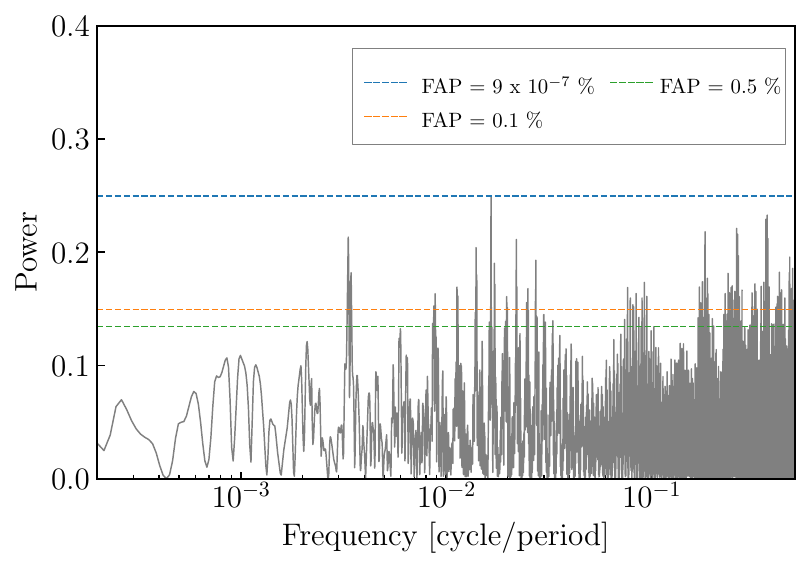} 
    \caption{The GLS periodograms were calculated for the timing residuals of a total of 188 epochs. The dashed line indicates the FAP levels. The highest power peak was found at a frequency of 0.01657 $\pm$ 0.00001 cycles/day, with a FAP of $9 \times 10^{-7} $\%.}
    \label{fig:gls_all}
\end{figure*}

In order to investigate the periodicity of the timing residuals ($O-C$) data for WASP-43~b, shown in \Cref{tab:timing}, we used the Generalized Lomb-Scargle periodogram (GLS; \citealt{zech2009}) in the {\tt\string PyAstronomy}\footnote{PyAstronomy: \texttt{https://github.com/sczesla/PyAstronomy}} routines \citep{zesla2019}. The GLS analysis was performed on all 188 timing residual ($O-C$) data points. The periodogram showed the highest power peak of 0.25 at a frequency of 0.01657 $\pm$ 0.00001 cycles/day, which corresponds to a False Alarm Probability (FAP) of $9 \times 10^{-7} $\% (\Cref{fig:gls_all}). Given the high obtained FAP values, no significant TTV signals were detected in our analysis.

\begin{figure*}[htb]
\centering
    \includegraphics[width=0.45\textwidth,page=1]{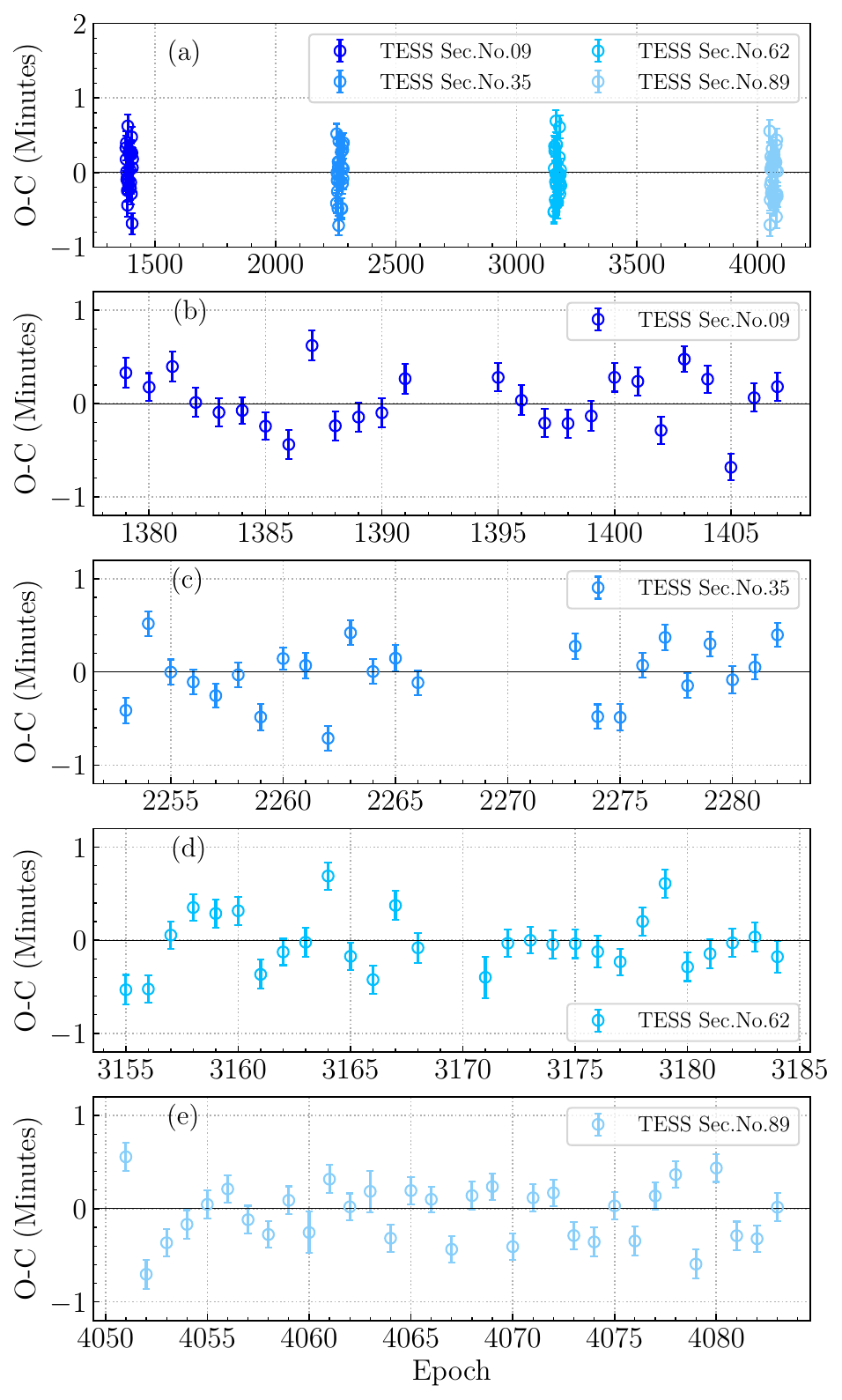} 
    \caption{The $O-C$ diagram for the \emph{TESS} data sets. (a) The $O-C$ data from observations in \emph{TESS} three sectors over a five-year interval. (b)-(e) The $O-C$ data plotted by individual sectors for clearer visibility.}
    \label{fig:oc_tess}
\end{figure*}

\begin{figure*}[htb]
\centering
    \includegraphics[width=0.45\textwidth,page=1]{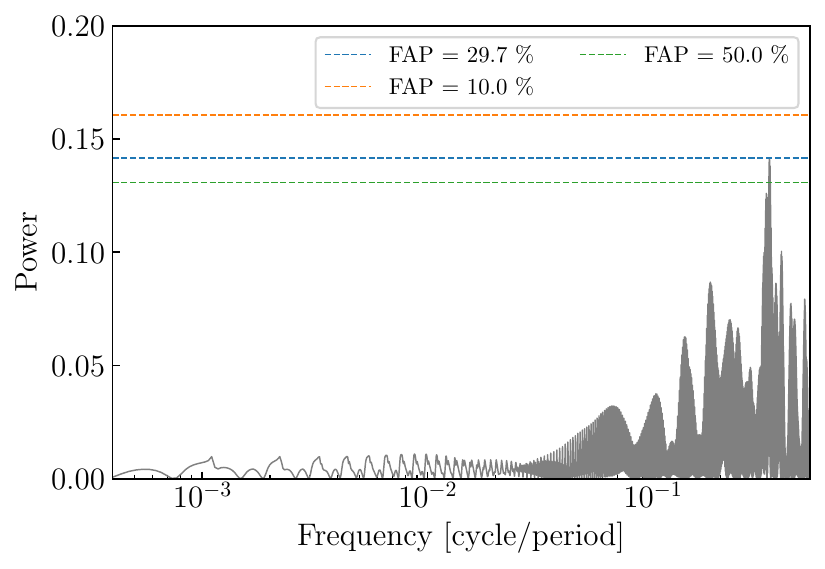} 
    \caption{The GLS periodograms of the timing residuals for \emph{TESS} data show that the highest power peak was found at a frequency of 0.33017 $\pm$ 0.00005 cycles/day with a False Alarm Probability (FAP) of 29.7\%.}
    \label{fig:gls_tess}
\end{figure*}

The \emph{TESS} transits in Sectors 09 and 35 were investigated for a downward trend in the orbital period by \citet{davoudi2021}. They found a decrease in the orbital period of $(-1.15 \pm 0.76) \times 10^{-7}$ in the slope line between a two-year interval of \emph{TESS} observations, suggesting that the target is more interesting for follow-up observations. Motivated by this, we performed a timing variation analysis for these \emph{TESS} timing residual ($O-C$) data, including the latest \emph{TESS} data from Sectors 62 and 89. The $O-C$ data derived for \emph{TESS} data are shown in \Cref{fig:oc_tess}. Similar to the total of 188 $O-C$ data points, the $O-C$ of the \emph{TESS} data was searched for timing variability over a five-year interval. Using GLS, the periodogram shows the highest power peak of 0.142 at a frequency of 0.33017 $\pm$ 0.00005 cycles/period (epoch) with an FAP of 29.7\% (\Cref{fig:gls_tess}).
    
\subsection{Upper-mass Limit for an Additional Planet}
Since no TTV signal was detected from the timing data, the 188 mid-transit times from \texttt{TransitFit} were used to simulate the upper mass limit of a potential nearby planetary companion. To estimate the upper mass limit of the second planet, we adapted the method described by \citet{awiphan2016,athano2022}. The orbit of the second planet is assumed to be circular and coplanar with the orbit of WASP-43~b. We calculated the unstable regions from the mutual Hill sphere between WASP-43~b and the perturbing planet, as described by \citet{fab2012}.;
\begin{equation}
    r_{H} = \frac{a_{\textup{in}} + a_{\textup{out}}}{2}\left(\frac{M_{\textup{in}} + M_{\textup{out}}}{3M_\star}\right)^{1/3} \ ,
\end{equation}
where $a_{\textup{in}}$ and $a_{\textup{out}}$ are the semi-major axis of the inner and outer planets (perturber planet), respectively. The boundary of stable orbit is when the separation of the planets semimajor axes ($a_{\textup{out}} - a_{\textup{in}}$) is larger than $2\sqrt{3}$ of the mutual Hill sphere. 

\begin{figure*}[htb]
\centering
    \includegraphics[width=0.6\textwidth,page=1]{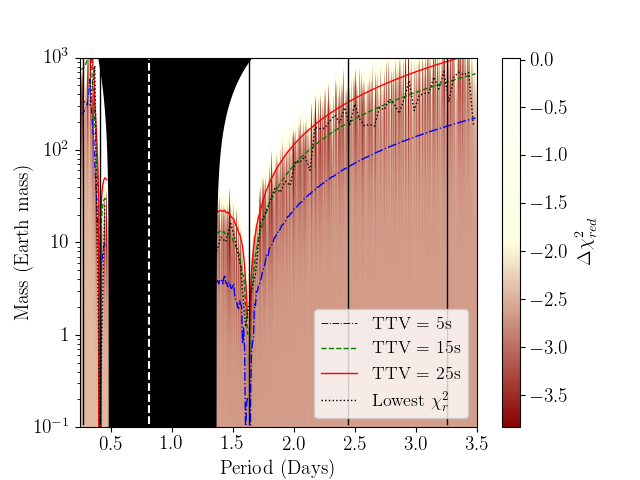} 
    \caption{Upper mass limit of the perturbing planet in the WASP-43 system. The best $\Delta{\chi}^{2}_{\textup{red}}$ values, binned with a period ratio of 0.05, are shown as the black dotted line. The upper mass limits for TTV amplitudes of 5, 15, and 25 seconds are represented by the blue dash-dotted, green dashed, and red solid lines, respectively. The white vertical line represents the orbital period of WASP-43~b. The black vertical lines, from left to right, indicate the orbital period resonances of 2:1, 1:2, 1:3, and 1:4. The contour illustrates the $\Delta{\chi}^{2}_{\textup{red}}$ values from comparing the best TTV model to the best linear model. The unstable orbit region is shown as the shaded black area.}
    \label{fig:upperma}
\end{figure*}

The TTV signal was computed using the \texttt{TTVFaster} package\footnote{TTVFaster: \texttt{https://github.com/ericagol/TTVFaster}}, developed by \citet{deck2016}, a tool for dynamical analysis. To obtain the accurate value of the TTV signal from \texttt{TTVFaster}, we employed the dynamic nested sampling algorithm from \texttt{dynesty} \citep{speagle2020},  using 200 live points and stopping when $\Delta \log Z = 0.5$. The period ratio between the second planet and WASP-43~b was varied from 0.3 to 4.50 with a step size of 0.01, while the mass range was set from $10^{-1}M_{\oplus}$ to $10^{3}M_{\oplus}$ on a logarithmic scale. The initial phase of the second planet was allowed to vary between 0 and $2\pi$. From the $O-C$ diagram in \Cref{fig:oc}, a signal amplitude of 15.40 seconds was found. We also calculated the upper mass limits corresponding to TTV amplitudes of 5, 15, and 25 seconds, as shown in \Cref{fig:upperma}.

Then, we calculated the difference of chi-square value  ($\Delta{\chi}^{2}_{\textup{red}}$) by comparing the signal from the TTV model, representing the best fit for the two-planet model, $\chi^{2}_{\textup{red}}$, to the best-fit single-planet model or linear fitting model, with $ \chi^2_{red,l}$ of 18.8. In \Cref{fig:upperma}, the $\Delta{\chi}^{2}_{\textup{red}}$ is plotted as a function of the perturbing mass and period, showing that the lowest $\Delta{\chi}^{2}_{\textup{red}}$ values are close to the upper mass limit corresponding to a TTV amplitude of 15 seconds. In the unstable orbit region with an orbital period ranging from 0.49 to 1.36 days, the presence of a second nearby planet is excluded. Based on the simulation, we can conclude that no planet with a mass heavier than $10^{2}M_{\oplus}$ exists with a period of less than two days.

\section{WASP-43 b's planetary atmosphere } 
\label{sec:atmosphere}
The transit depth data ranging from optical to mid-infrared wavelengths, derived from the light curve fitting in Section~\ref{sec:LCModeling}, is used to re-examine the chemical compositions of WASP-43~b atmosphere. In previous studies, the transmission spectrum of WASP-43~b obtained from \emph{HST}/WFC3 G141 was utilized to determine the atmospheric composition, with the presence of H$_{2}$O detected by \citet{kreidberg2018,tsiaras2018,weaver2020}. Furthermore, \citet{chubb2020} analyzed the transmission spectrum provided by \citet{kreidberg2018} and reported a highly significant detection of AlO, with a confidence level exceeding $5\sigma$ compared to a flat baseline model. Therefore, we first focus on the transit depth data from \emph{HST}/WFC3 G141 to compare the results obtained from \texttt{TransitFit} with those from previous studies.

To retrieve the transmission spectrum of WASP-43~b, the open-source Bayesian atmospheric retrieval framework \texttt{TauREx3} \citep{al-rafaie2021}\footnote{\texttt{TauREx3}: \texttt{https://github.com/ucl-exoplanets/TauREx3\_public/}}, which uses the nested sampling routines from \texttt{MultiNest} \citep{feroz2009} with 1000 live points to determine the atmospheric parameters. For the transmission spectrum modeling with the \texttt{TauREx3} package, the stellar parameters and planetary mass of the WASP-43 host star were taken from \citet{esposito2017}. The stellar spectrum for the host star, with a temperature of $T_* = 4166$ K, was simulated using a \texttt{PHOENIX} model \citep{husser2013}. An isothermal temperature profile was assumed, using a plane-parallel atmosphere consisting of 100 layers. The cloud-top pressure was allowed to vary from $10^1$ to $10^6$ Pa on a logarithmic scale. A He/H$_2$ ratio of 0.17, consistent with the solar abundance, was used. The transmission spectra were generated at a resolution of 10,000 before being binned to match the instrumental resolution.

Following the initial study of the transmission spectrum of WASP-43~b by \citet{kreidberg2018}, we modeled the transmission spectrum by considering molecular opacities as described by \citet{kreidberg2014}. Specifically, H$_2$O \citep{polyansky2018}, CH$_4$ \citep{yurchenko2024}, CO \citep{chubb2021}, CO$_2$ \citep{yurchenko2020}, and AlO \citep{patrascu2014} were included. Additionally, we added potential metal oxides such as TiO \citep{mckemmish2019} and VO \citep{mckemmish2016} which have spectral features in the visible waveband in the model. The molecular line lists were obtained from the ExoMol \citep{tennyson2016}, HITRAN \citep{gordon2016}, and HITEMP \citep{rothman2014} databases. We also accounted for collision-induced absorption between H$_2$ molecules \citep{abel2011,fletcher2018} and between H$_2$ and He \citep{abel2012}. A list of priors for the parameters and chemical abundances used in the \texttt{TauREx3} retrieval is provided in \Cref{tab:atmosphere_para}.

Previous studies have utilized \texttt{TauREx3} to retrieve the atmospheric compositions of WASP-43~b using transmission spectra obtained via the \texttt{Iraclis} package \citep{tsiaras2016}, which uses \texttt{PyLightcurve} \citep{tsiaras2016-py} to fit the white light curve and correct systematics before fitting the spectral light curves. Adopting this framework, we analyzed the \emph{HST}/WFC3 G141 data using the \texttt{Iraclis} package (Case I) and compared the results with transit depths derived using \texttt{TransitFit} (Case II). Furthermore, for the comprehensive atmospheric retrieval analysis, we combined these \emph{HST}/WFC3 G141 datasets with transit depths in other wavebands derived from \texttt{TransitFit}, defining the combined \texttt{Iraclis}-based dataset as Case III and the combined \texttt{TransitFit}-based dataset as Case IV.

\begin{figure*}[htb]
\centering
    \includegraphics[width=0.45\textwidth,page=1]{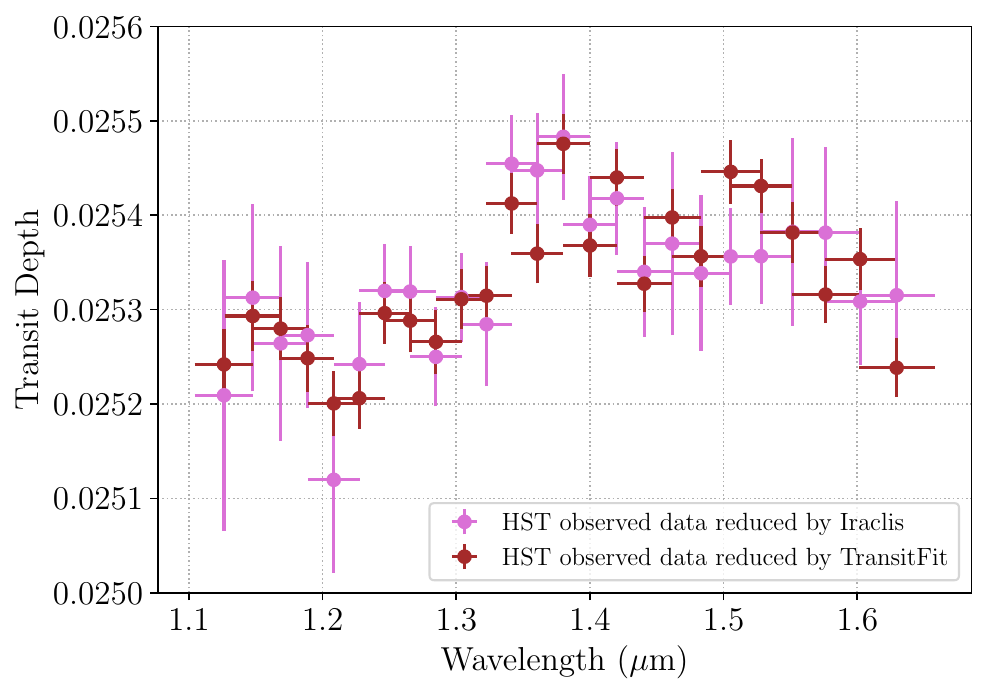} 
    \caption{The \emph{HST} transmission spectrum of WASP-43~b, reduced using \texttt{Iraclis} (pink) and \texttt{TransitFit} (dark-red).}
    \label{fig:hst-raw}
\end{figure*}

For the spectra obtained via the \texttt{Iraclis} package, we utilized the best-fit values of planetary parameters from \Cref{tab:outpara}. We analyzed both the \texttt{Iraclis} and \texttt{TransitFit} reductions to investigate the observed discrepancy in their transit depths, as shown in \Cref{fig:hst-raw}. We determined that this difference arises primarily from the treatment of Limb Darkening Coefficients (LDCs). Specifically, \texttt{Iraclis} utilizes LDCs from \citet{claret2000}, whereas \texttt{TransitFit} employs LDCs derived from \texttt{ExoCTK}. Given that the choice of LDC models has a significant impact on the retrieved transit depths, we decided to include the analysis of both reductions to ensure a comprehensive comparison.

\subsection{Case I: \emph{HST}/WFC3 G141 Transmission Spectrum from \texttt{Iraclis}}

For the fitting of the \emph{HST}/WFC3 G141 transit light curves of WASP-43~b, we assumed a circular orbit. The orbital period ($P$), inclination ($i$), and semimajor axis ($a$) were taken from \Cref{tab:outpara} and, along with the mid-transit time ($t_{0} = 2457423.44976$ BJD$_{\textup{TDB}}$), were fixed. The limb-darkening coefficients were modeled using the non-linear four-coefficient law from \citet{claret2000}. The host-star parameters were defined consistently with the \texttt{TransitFit} light-curve analysis, adopting an effective temperature of $T = 4166 \pm 100$ K and metallicity $Z{} = -0.05 \pm 0.17$ from \citet{bonomo2017}, and $\log(g_{}) = 4.6 \pm 0.1$ from the Gaia EDR3 catalog. The planet-to-star radius ratio ($R_p/R_*$) was allowed to vary during the fitting process. The transit depths computed by \texttt{Iraclis} are provided in \Cref{tab:Rp_Iraclis}.

From the posterior retrieval of Case I, we obtained a planetary temperature of $900^{+500}_{-400}$ K, consistent with previous studies by \citet{tsiaras2018} ($957.27 \pm 343.3$ K) and \citet{chubb2020} ($857.66^{+419.17}_{-288.89}$ K). The retrieved log water volume mixing ratio was $-0.5^{+0.3}_{-2.3}$, with a reduced chi-square of $\chi^{2}_{\textup{red}} = 1.70$. However, the posterior distribution exhibited two distinct solutions (see \Cref{fig:contour-HST}). We therefore applied a k-means clustering algorithm to separate the results into two sub-cases.

In Case I.I, the derived temperature was $500^{+200}_{-100}$ K, consistent within $1\sigma$ of the value reported by \citet{kreidberg2014} ($639.5^{+144.6}_{-129.2}$ K). The retrieved log water abundance was $-2^{+1}_{-2}$, consistent within $1\sigma$ with \citet{edwards2003,tsiaras2016,weaver2020,chubb2020,bartelt2025}. In Case I.II, the temperature was $1200^{+300}_{-300}$ K, which agrees within $1\sigma$ with the study by \citet{bartelt2025} ($1013.17^{+147.45}_{-119.45}$ K) and \citet{tsiaras2018}. The retrieved log water abundance was $-0.4^{+0.2}_{-0.1}$ for Case~I.II, consistent with the value found in the overall Case I retrieval. The $\chi^{2}_{\textup{red}}$ values for Case I.I and Case I.II were 1.95 and 1.39, respectively. The three retrievals of the \emph{HST}/WFC3 G141 transmission spectrum reduced with \texttt{Iraclis} showed no additional molecular species were detected at a significant level in any of the cases. The best-fit transmission spectrum for the HST data processed with \texttt{Iraclis} is displayed in \Cref{fig:HST-Spectrum}, and the resulting atmospheric parameters are listed in \Cref{tab:atmosphere_para}.

\subsection{Case II: \emph{HST}/WFC3 G141 Transmission Spectrum with \texttt{TransitFit}}

For the \emph{HST}/WFC3 G141 light curves fitted with \texttt{TransitFit}, the retrieved planetary temperature was $1200^{+200}_{-100}$\,K, matching the result from Case~I.II and consistent with \citet{tsiaras2018,bartelt2025}. The log water abundance was $-0.5^{+0.1}_{-0.1}$ with a $\chi^{2}_{\text{red}} = 5.41$. While this value is notably higher than the $\chi^{2}_{\text{red}}$ of 1.7 obtained in Case I, the difference is primarily driven by the transit depth errors from \texttt{TransitFit} being approximately half the size of those from \texttt{Iraclis}. This difference in error estimation is a result of the different sampling algorithms used, specifically Nested Sampling (\texttt{dynesty}) in \texttt{TransitFit} and MCMC (\texttt{emcee}) in \texttt{Iraclis}. Furthermore, variations in the detrending and limb darkening methods also contribute to this discrepancy. Despite the different $\chi^{2}_{\text{red}}$ values, the two fits are statistically consistent. Aside from the water feature discussed, no other molecular species were significantly detected.

\begin{figure*}[htb]
\centering
    \includegraphics[width=0.5\textwidth,page=1]{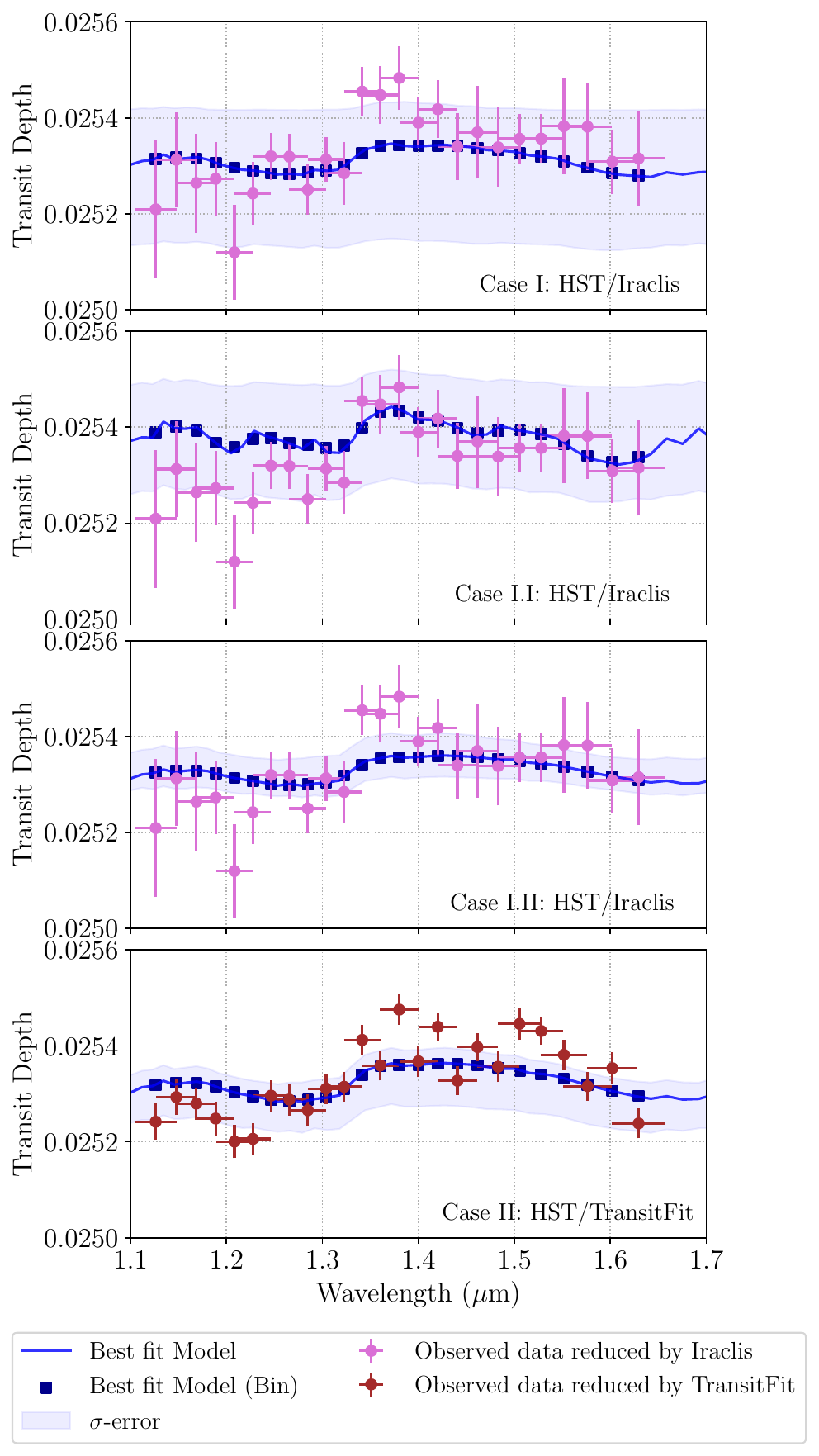} 
    \caption{The best-fit transmission spectrum model of WASP-43~b derived from \emph{HST} data. The synthetic model generated by \texttt{TauREx} is shown as a solid blue line, with the corresponding $1\sigma$ confidence region indicated by the blue shading. The binned best-fit model values are shown as blue squares. The transit depths computed with \texttt{Iraclis} and \texttt{TransitFit} are shown as pink and dark-red dots, respectively.} 
    \label{fig:HST-Spectrum}
\end{figure*}

The retrieval results are summarized in \Cref{tab:atmosphere_para}, with the best-fit spectrum shown in \Cref{fig:HST-Spectrum,fig:contour-HST}. From the analysis of the \emph{HST}/WFC3 G141 transmission spectra reduced with both \texttt{Iraclis} and \texttt{TransitFit}, the retrieved water abundances were consistently associated with the higher-temperature solutions, as seen in Case~I, Case~I.II, and the \texttt{TransitFit} retrieval (Case~II).

\subsection{Case III: Full Transmission Spectrum using \texttt{Iraclis}-reduced \emph{HST} Data}

As with Case~I and Case~II, our initial analysis was limited to transmission spectra obtained from \emph{HST}/WFC3 G141. To achieve a comprehensive statistical analysis of the atmospheric chemical composition of WASP-43~b, we expanded the dataset by combining the \emph{HST} spectra with observations from ground-based facilities, \emph{TESS}, and \emph{JWST}. This extension broadens the wavelength coverage from the 1.1--1.6\,$\mu$m range to 0.3--10\,$\mu$m. Given that the \emph{HST} transmission spectra in Case~I and Case~II were processed using different reduction tools, we maintained this distinction in the combined analysis, separating them into Case~III and Case~IV.

For the \emph{JWST}/MIRI data, we utilized the spectroscopic light curves directly from \citet{bell2024}. We applied a second-order polynomial detrending model within \texttt{TransitFit} to independently determine the transit depth for each channel. Unlike the discrepancy observed in the \emph{HST} data analysis, the resulting transmission spectrum derived via \texttt{TransitFit} for the \emph{JWST} data is consistent with the values reported by \citet{bell2024} using the \texttt{Eureka!} pipeline.

In total, we utilized 82 transmission spectra, excluding the Clear filter data listed in \Cref{tab:radius-transitDepth-limbdark}. We note that an instrumental offset was applied to the \emph{JWST}/MIRI observations relative to the other datasets, with the magnitude of this offset determined via a weighted average of the combined data, as in \citet{grant2023}. Additionally, while the standard \texttt{PHOENIX} stellar models cover the spectral range from 50\,nm to 5.5\,$\mu$m, the \emph{JWST}/MIRI spectrum extends from 5 to 12\,$\mu$m. To address this, we extrapolated the \texttt{PHOENIX} model to fit the transmission spectra beyond 5.5\,$\mu$m.

\begin{figure*}[htb]
\centering
    \includegraphics[width=0.85\textwidth,page=1]{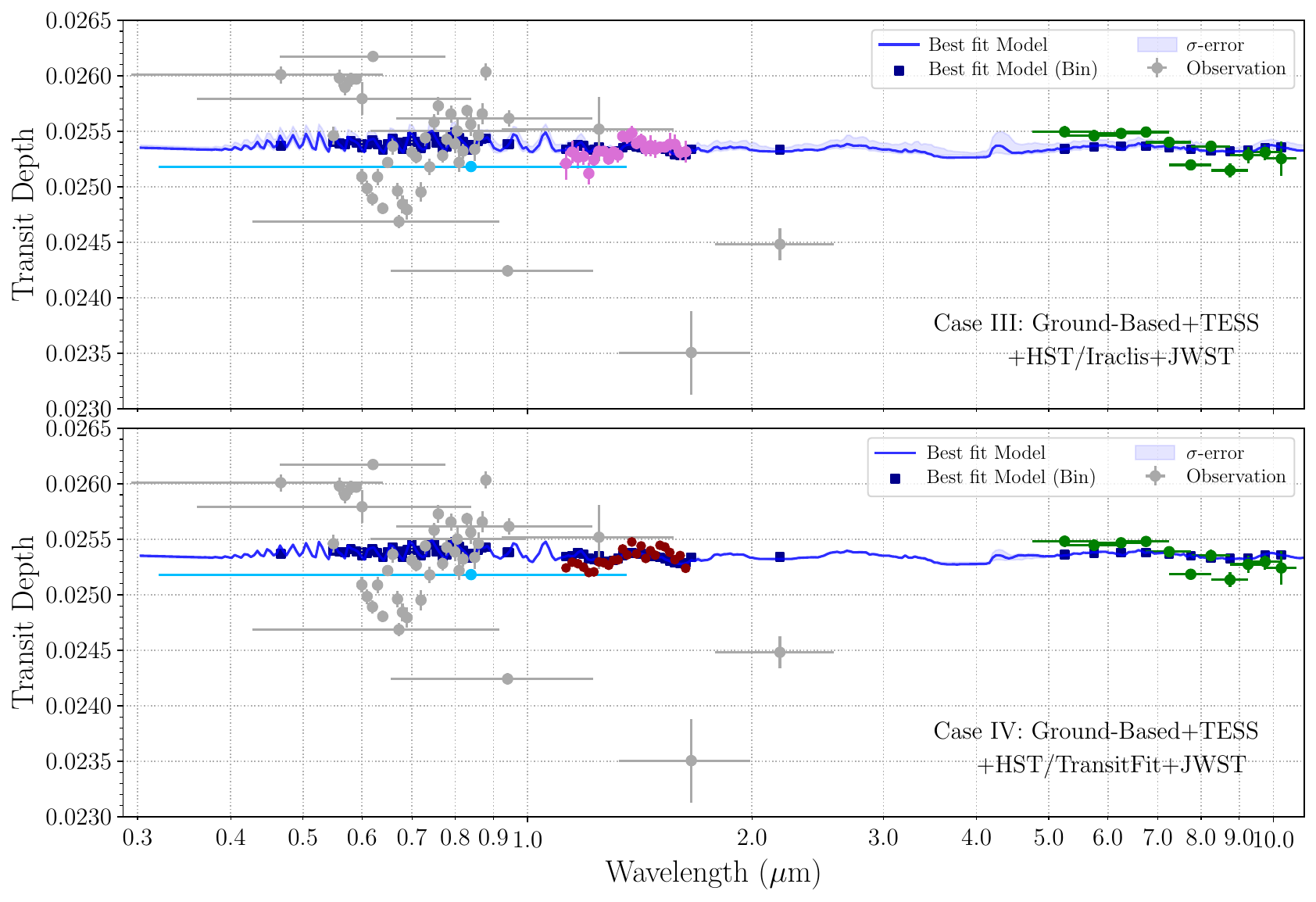} 
    \caption{The best-fit transmission spectrum model of WASP-43~b, calculated for the combined dataset including ground-based observations (grey dots), \emph{TESS} (light-blue dots), \emph{HST} processed with \texttt{Iraclis} (pink dots), \emph{HST} processed with \texttt{TransitFit} (dark-red dots), and \emph{JWST} (green dots). The synthetic model generated by \texttt{TauREx} is shown as a solid blue line, with the corresponding $1\sigma$ confidence region indicated by the blue shading. The binned best-fit model values are shown as blue squares. Note that the five outlier transit depths in the optical wavelength range correspond to the $R$, $I$, $z'$, $H$, and $K$ filters, respectively.} 
    \label{fig:GHTJ-Spectrum}
\end{figure*}

The best-fit transmission spectrum for Case~III is shown in \Cref{fig:GHTJ-Spectrum}, with the corresponding posterior distribution in \Cref{fig:contour-GTHJ}. Despite the inclusion of an instrumental offset for the \emph{JWST}/MIRI spectrum, the combination of data over such a broad wavelength range introduced significant modeling complexities, resulting in high residuals that make the retrieval of reliable atmospheric parameters difficult. We find that in this regime, the evidence for H$_2$O diminishes to non-significant levels, and no other molecular species are robustly detected.

\subsection{Case IV: Full Transmission Spectrum using \texttt{TransitFit}-reduced \emph{HST} Data}

We define Case~IV by combining the \emph{HST}/WFC3 G141 transmission spectra reduced with \texttt{TransitFit} (Case~II) with the transmission spectra from ground-based observations, \emph{TESS}, and \emph{JWST} which were also reduced with \texttt{TransitFit}. Similar to Case~III, the increased complexity of the broad-baseline retrieval for Case~IV limited our ability to achieve a statistically robust model fit. This confirms that when the full wavelength coverage is considered with current data, the water abundance is not detected at a significant level. High-precision data across these wavelengths remains essential to overcome these complexities and break existing model degeneracies.

\begin{table*}
\begin{center}
\caption{Parameters and their priors used in \texttt{TauREx 3} retrieval, along with the best-fit retrieved values with  $1\sigma$ uncertainties for six case studies.}
\label{tab:atmosphere_para}          
\begin{tabular}{lcccccccc}
\toprule
\multirow{2}{*}{Parameter}  & \multirow{2}{*}{Priors}  & \multirow{2}{*}{Scale} & \multicolumn{6}{c}{Retrieved Values}   \\
\cline{4-9}
  & & & Case I &  Case I.I &  Case I.II  & Case II & Case III & Case IV \\  
\hline
$T$ (K)    & (200, 2000) & linear & $900^{+500}_{-400}$ & $500^{+200}_{-100}$ & $1200^{+300}_{-300}$  &  $1200^{+200}_{-100}$ & $400^{+100}_{-20}$ & $400^{+20}_{-40}$\\
H$_{2}$O   & (-4.0, -1.0)  & log    & $-0.5^{+0.3}_{-2.3}$ & $-2^{+1}_{-2}$ & $-0.4^{+0.1}_{-0.2}$  & $-0.5^{+0.1}_{-0.1}$ & $-2.8^{+0.5}_{-1.2}$ & $-2.1^{+0.3}_{-0.4}$ \\
CH$_{4}$   & (-9.0, -1.0)  & log     & $-6^{+3}_{-3}$ & $-5^{+2}_{-3}$ & $-6^{+3}_{-3}$  &   $-6^{+3}_{-2}$ & $-7^{+2}_{-2}$ & $-8^{+2}_{-1}$  \\
CO         & (-9.0, -1.0)  & log     & $-6^{+3}_{-3}$ & $-6^{+3}_{-3}$ & $-6^{+3}_{-3}$ &  $-6^{+3}_{-3}$ & $-7^{+2}_{-2}$ & $-4^{+1}_{-3}$ \\
CO$_{2}$   & (-9.0, -1.0)  & log     & $-6^{+3}_{-3}$ & $-6^{+3}_{-3}$ & $-6^{+3}_{-3}$ &   $-6^{+3}_{-3}$ & $-6^{+2}_{-3}$ & $-6^{+2}_{-2}$ \\
TiO        & (-9.0, -1.0)  & log     & $-6^{+3}_{-3}$ & $-6^{+2}_{-3}$ & $-6^{+3}_{-3}$ &  $-6^{+2}_{-2}$  & $-9.5^{+0.7}_{-0.4}$ & $-9.5^{+0.7}_{-0.4}$ \\
VO         & (-9.0, -1.0)  & log   & $-8^{+2}_{-1}$ & $-8^{+2}_{-1}$ & $-8^{+2}_{-2}$ & $-8^{+1}_{-1}$  & $-6.2^{+0.4}_{-1.1}$ & $-6.1^{+0.4}_{-0.4}$ \\
AlO         & (-9.0, -1.0)  & log   & $-6^{+2}_{-2}$ & $-6^{+2}_{-2}$ & $-7^{+2}_{-2}$ & $-7^{+2}_{-2}$ & $-6.6^{+0.7}_{-0.8}$ & $-8^{+1}_{-1}$  \\
$\log(P_{\text{clouds}})$ (Pa)   & ($10^{1}$, $10^{6}$)  & log   & $4^{+1}_{-2}$ & $4^{+1}_{-1}$ & $3^{+2}_{-2}$ & $2.7^{+2.0}_{-0.8}$ & $5.1^{+0.5}_{-0.6}$ & $-5^{+1}_{-1}$ \\

\hline
\multicolumn{3}{c}{$\chi^{2}_{\textup{red}}$} & 1.70  & 1.95 & 1.39 & 5.41 & 34.35 & 34.29 \\
\hline
\end{tabular}
\end{center}
\end{table*}


\section{Conclusions}
\label{sec:conclude}

In this study, we conducted ground-based multi-band follow-up observations of the Hot Jupiter WASP-43~b using the SPEARNET network. A total of 35 transit light curves were obtained and combined with data from \emph{HST}, \emph{TESS}, \emph{JWST}, and 109 previously published ground-based light curves. These datasets were used to model the transit light curves and refine the planetary parameters of WASP-43~b using \texttt{TransitFit}.  

A total of 188 mid-transit times, measured with \texttt{TransitFit}, were analyzed to investigate potential timing variations in WASP-43~b. No strong evidence for orbital decay was found, with $\textup{d}P_{d}/\textup{d}E = -3^{+1}_{-1} \times 10^{-11}$ days per orbit. Additionally, periodogram analysis also showed no significant transit timing variations (TTVs). While \citet{davoudi2021} reported a downward trend in \emph{TESS} data from Sectors 09 and 35, our analysis of timing residuals ($O-C$) from four \emph{TESS} sectors, including the latest Sector 89, showed no variability over a seven-year interval. Moreover, simulations using \texttt{TTVFaster} further suggest that no planet with a mass greater than $10^{2}M_{\oplus}$ could exist with an orbital period of less than two days.  

The atmospheric retrieval analysis of WASP-43~b reveals a distinct dependency of the retrieved parameters on wavelength coverage and data reduction methodology. While the \emph{HST}-only retrievals with both \texttt{Iraclis} (Case~I) and \texttt{TransitFit} (Case~II) indicate planetary temperatures ranging from 500~K to 1200~K with higher water abundances, aligning with previous studies. The combined broader optical-to-mid-infrared baseline from SPEARNET, \emph{TESS}, and \emph{JWST} (Case~III and Case~IV) presents significant modeling challenges that limit the reliability of atmospheric characterization. In these two specific cases, the integration of data across such an extensive wavelength range introduced substantial complexities, making it difficult to achieve a statistically robust model fit. Therefore, high-precision transmission spectroscopy spanning a broad wavelength range is essential to overcome these modeling complexities and achieve a more reliable atmospheric characterization.

\vspace{6mm}
We thank the referee for the helpful and constructive comments that have improved this work. The observation data used in this work based on observations made with ULTRASPEC at the Thai National Observatory, the Thai Robotic Telescopes, and the Regional Observatories for the Public under the operation of the National Astronomical Research Institute of Thailand (Public Organization). The simulation section in this work was performed using the CHALAWAN NARIT High-Performance Computing (CHALAWAN HPC) system.

This work used the available data based on observations made with the NASA/ESA \emph{Hubble Space Telescope} (\emph{HST}) obtained from the Space Telescope Science Institute (STScI), which is operated by the Association of Universities for Research in Astronomy (AURA), Inc., under NASA contract NAS 5–26555. The published \emph{HST} data present here were taken as part of proposal 13467, led by Jacob Bean. This paper also includes data collected with the \emph{TESS} mission, obtained from the MAST data archive at the STScI. Funding for the \emph{TESS} mission is provided by the NASA Explorer Program. STScI is operated by the AURA, Inc., under NASA contract NAS 5–26555. The specific observations analyzed can be accessed via \dataset[{https://doi.org/10.17909/T97P46}]{} and \dataset[{https://doi.org/10.17909/t9-nmc8-f686}]{}, respectively. This work also used the data based on observations made with the NASA/ESA/CSA James Webb Space Telescope. The data were obtained from the MAST at the STScI, which is operated by the AURA, Inc., under NASA contract NAS 5-03127 for \emph{JWST}. These observations are associated with program proposal JWST-ERS-1366, led by Taylor J. Bell.  This research made use of the open source Python package \texttt{ExoCTK}, the Exoplanet Characterization Toolkit \citet{ExoCTK}. 

We thank Zoltán Garai for providing the observational data from MuSCAT2 and Quentin Changeat for suggesting the instructions on TauREx. This work is supported by the Fundamental Fund of Thailand Science Research and Innovation (TSRI) through the National Astronomical Research Institute of Thailand (Public Organization) (FFB680072/0269).

%

\vspace{5mm}

\facilities{\emph{HST}/WFC3 G141, \emph{TESS}, \emph{JWST}/MIRI, 2.4-m (TNT), 1-m (TNT), 0.5-m (TRT-TNO), 0.7-m (TRT-GAO), 0.7-m (TRT-SBO), 0.7-m (TRT-SRO), 0.7-m (ROP-NM) and 0.7-m (ROP-CC)}

\software{\texttt{sextractor} \citep{bertin1996}, \texttt{Astrometry.net} \citep{lang2010}, \texttt{TransitFit} \citep{hayes2024}, \texttt{TTVFaster} \citet{deck2016}, \texttt{Iraclis} \citep{tsiaras2016} and \texttt{TauREx} \citep{al-rafaie2021}.}



\appendix
\counterwithin{figure}{section}
\section{The planet radius and limb-darkening}
\begin{table*}
\begin{center}
\caption{The planet-to-star radius ratio ($R_p$/$R_{*}$), transit depth, and quadratic LDCs of WASP-43~b in different 83 filters, as obtained by \texttt{TransitFit}.}
\label{tab:radius-transitDepth-limbdark}          
\begin{tabular}{ccccccc}
\toprule
\multirow{2}{*}{Filter} & Mid-wavelength & Bandwidth & \multirow{2}{*}{ {$R_p$/$R_{*}$}} & \multirow{2}{*}{Transit Depth (\%)} & \multirow{2}{*}{$u_0$} & \multirow{2}{*}{$u_1$} \\ 
 & ($\mu$m) & ($\mu$m) &	 &  &	 &  \\ 
\hline				
$I+z'$-band	&	0.944	&	0.27	&	0.1600	$\pm$	0.0002	&	2.562	$\pm$	0.007	&	0.34	$\pm$ 0.01	& 0.33 $\pm$	0.03 \\
$r'$-band	&	0.621	&	0.16	&	0.1618	$\pm$	0.0001	&	2.617	$\pm$	0.004	&	0.570	$\pm$ 0.009	& 0.455 $\pm$	0.008 \\
$g'$-band	&	0.467	&	0.17	&	0.1613	$\pm$	0.0002	&	2.601	$\pm$	0.008	&	0.67	$\pm$ 0.02	& 0.56 $\pm$	0.02 \\
...	&	...	&	...	&	...	&	...	&	...	&	...	\\
...	&	...	&	...	&	...	&	...	&	...	&	...	\\
\hline
\end{tabular}
\end{center}
{\textbf{Notes.} The full table is available in a machine-readable format in the online journal.}  
\end{table*}

\begin{table*}
\begin{center}
\caption {The planet-to-star radius ratio ($R_p$/$R_{*}$) and their transit depths of WASP-43~b derived by \texttt{Iraclis}.}
\label{tab:Rp_Iraclis}
\begin{tabular}{ccccc}
\toprule
\multirow{2}{*}{Filter} & Mid-wavelength & Bandwidth & \multirow{2}{*}{ {$R_p$/$R_{*}$}} & \multirow{2}{*}{Transit Depth (\%)} \\ 
 & ($\mu$m) & ($\mu$m) &	 &   \\ 
\hline				
\emph{HST}/WFC3 G141 &	1.126 &	0.022&	0.1588	$\pm$0.0005 & 2.52	$\pm$ 0.01 \\
\emph{HST}/WFC3 G141 & 1.148 &	0.021&	0.1591	$\pm$0.0003 & 2.53	$\pm$ 0.01 \\
\emph{HST}/WFC3 G141 & 1.169 &	0.021&	0.1589	$\pm$0.0003 & 2.53	$\pm$ 0.01 \\
...	&	...	&	...	&	...	&	...	\\
...	&	...	&	...	&	...	&	...	\\
\hline
\end{tabular}
\end{center}
{\textbf{Notes.} The full table is available in a machine-readable format in the online journal.}  
\end{table*}

\section{WASP-43~b's Mid-transit Times and Timing Residuals}
\startlongtable
\begin{table*}
\begin{center}
\caption {Mid-transit Times ($t_{0}$) and Timing Residuals ($O-C$) for WASP-43~b from 188 epoch or transit event.}
\label{tab:midtransit}
\begin{tabular}{lccc}
\toprule
\multirow{2}{*}{Epoch} & $t_{0} +2400000$ & $(O-C)$ &   \multirow{2}{*}{Ref} \\
                       & (BJD$_{\textup{TDB}}$)    &  (days)       &     \\
\hline
-2318	&	55537.81672	$\pm$	0.00017	&	-0.00017	&	(a)	\\
-2307	&	55546.76491	$\pm$	0.00009	&	-0.00020	&	(a)	\\
-2302	&	55550.83219	$\pm$	0.00006	&	-0.00029	&	(a)	\\
...	&	...	&	...	&	...	\\
...	&	...	&	...	&	...	\\
\hline
\end{tabular}
\end{center}
{\textbf{Notes.} The full table is available in a machine-readable format in the online journal. Data Source: (a) \citet{gillon2012} (b) \citet{chen2014} (c) \citet{maciejewski2013}, (d) \citet{murgas2014}, (e) \emph{HST}, (f) \citet{ricci2015}, (g) \citet{jiang2016}, (h) This Study, (i) \citet{parvia2019} (j) \citet{murgas2014}, (k) \emph{TESS} and (l) \emph{JWST}.} 
\end{table*}

\section{The individual transit light curves from SPEARNET observations, \emph{HST} and \emph{TESS} modeled by \texttt{TransitFit} fitting.}

\begin{figure*}[htb]
\begin{tabular}{ll}
    \includegraphics[width=0.475\textwidth,page=1]{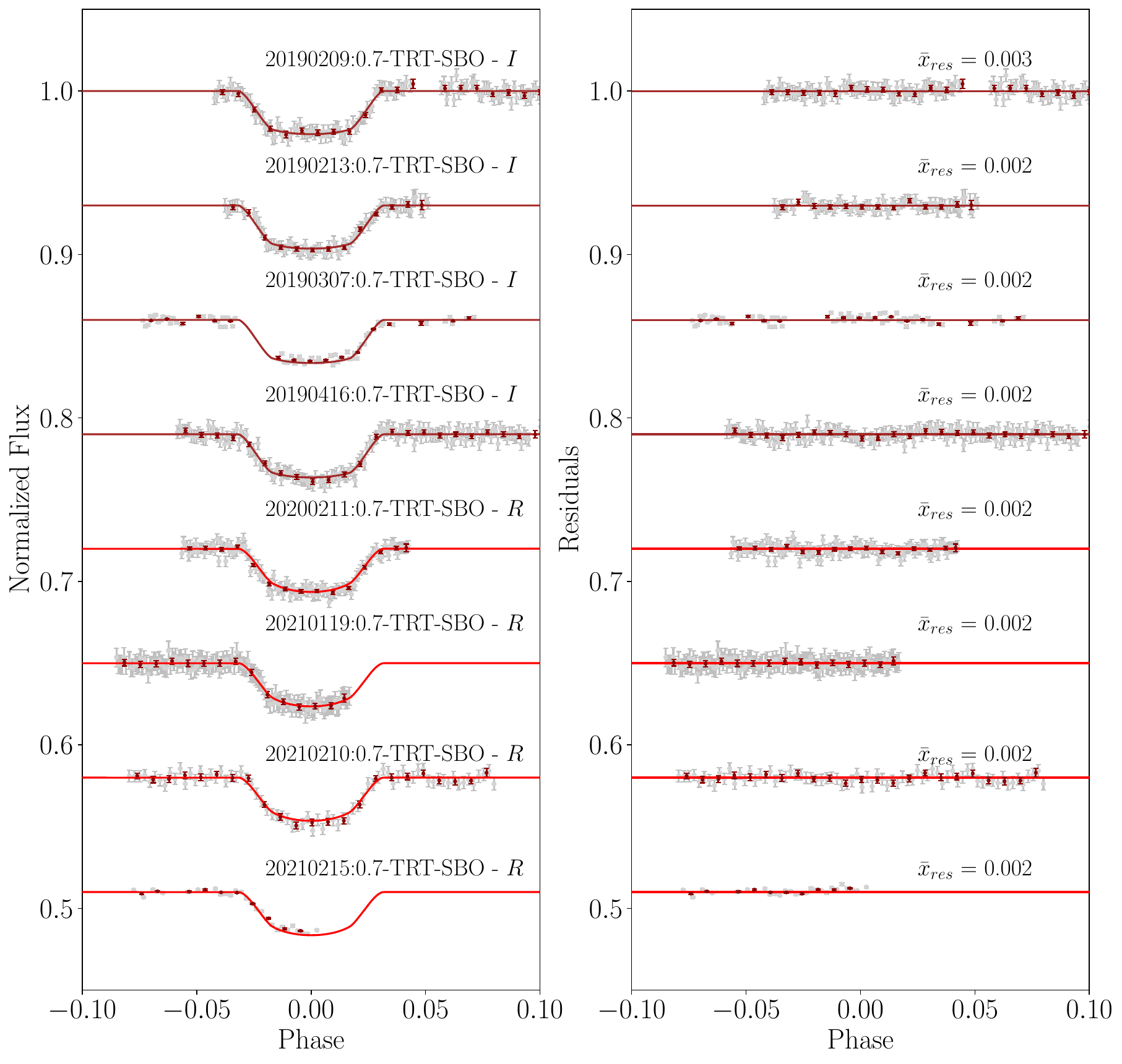} 
   \includegraphics[width=0.475\textwidth,page=1]{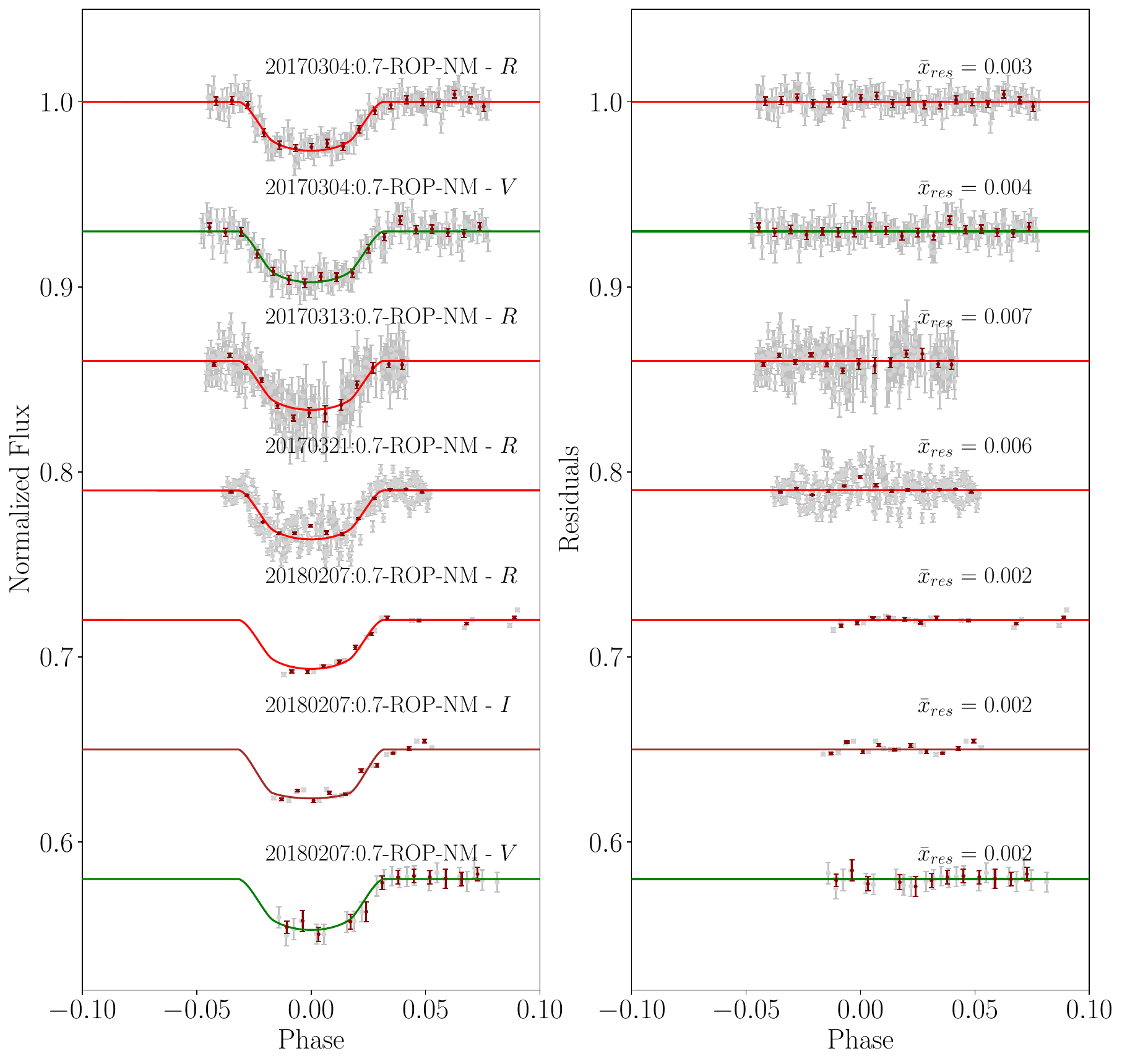} \\
    \includegraphics[width=0.475\textwidth,page=1]{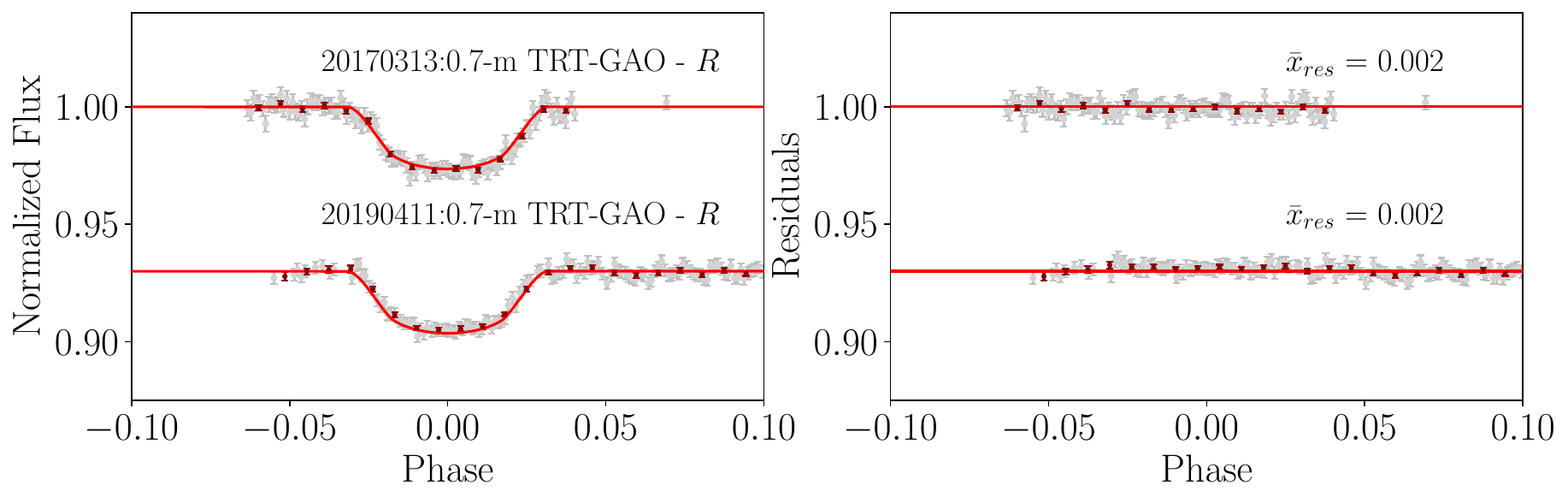} 
    \includegraphics[width=0.475\textwidth,page=1]{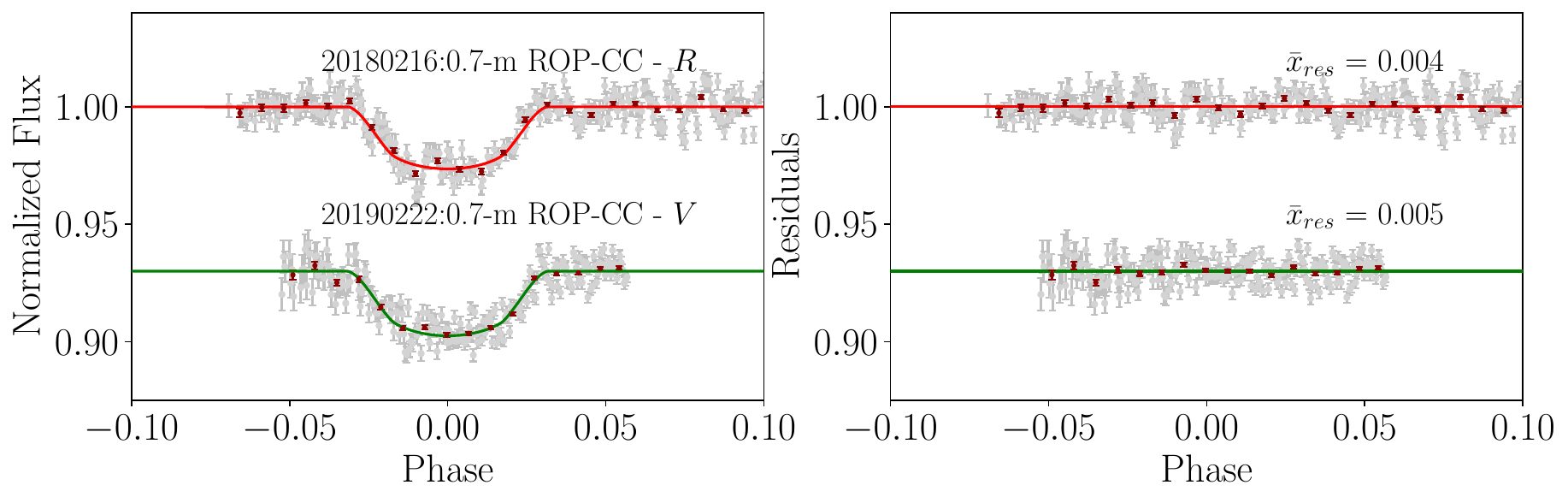} \\
   \includegraphics[width=0.475\textwidth,page=1]{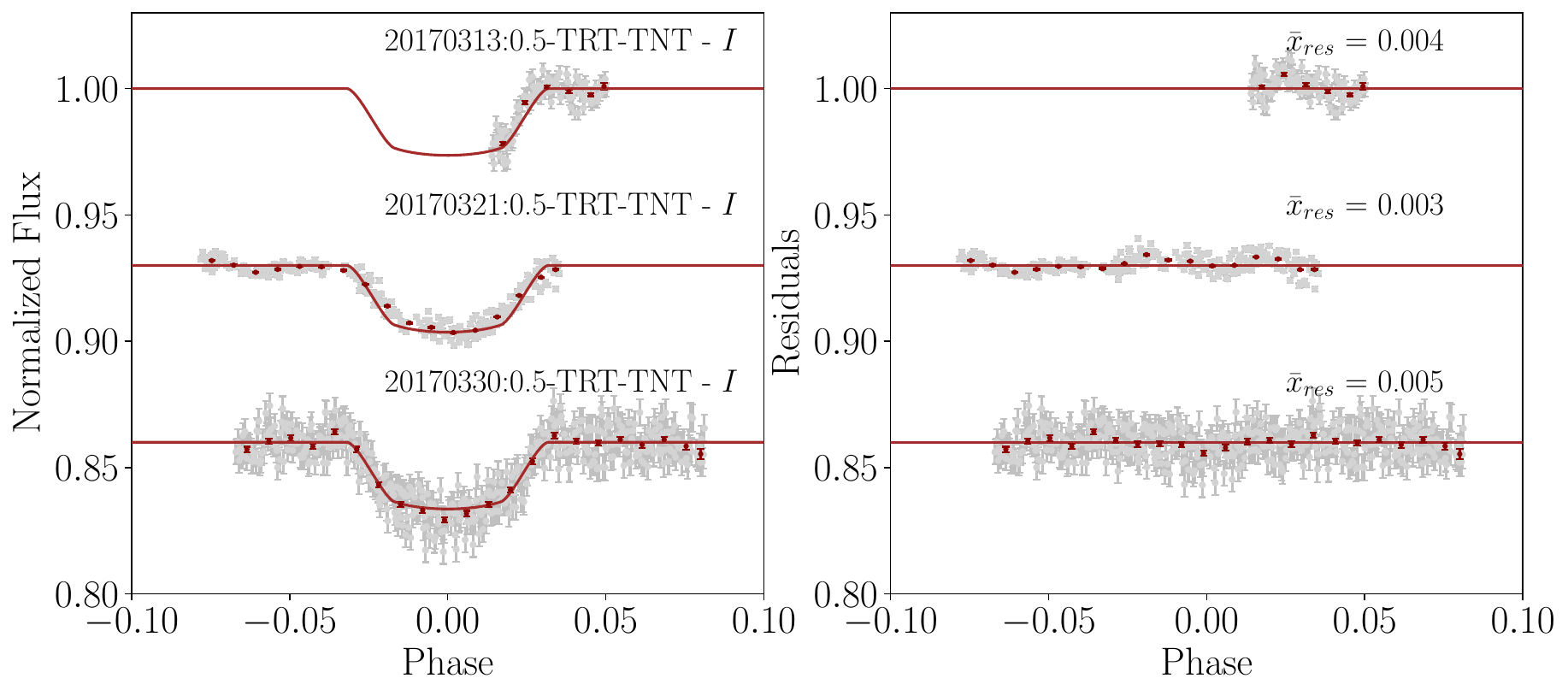} 
    \includegraphics[width=0.475\textwidth,page=1]{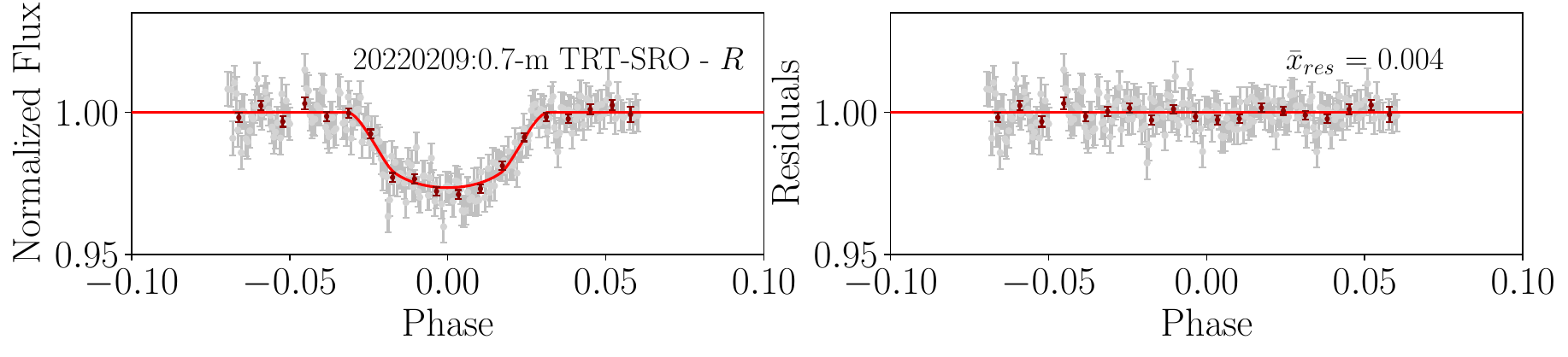} \\
   \end{tabular}
    \caption{The individual SPEARNET transit light curves of WASP-43~b observed at TRT-SBO, ROP-NM, TRT-GAO, ROP-CC, TRT-TNT, and TRT-SRO. The observed data are shown as gray dots, with filters $R$ (red), $I$ (brown), and $V$ (green). The corresponding residuals and the mean residual values (${\bar{x}_{res}}$) with clear offsets are displayed on the right panel.}
    \label{fig:LCs_individualTRTs}
\end{figure*}

\begin{figure*}[htb]
\centering
  \begin{tabular}{ccc}
    \includegraphics[width=0.325\textwidth,page=1]{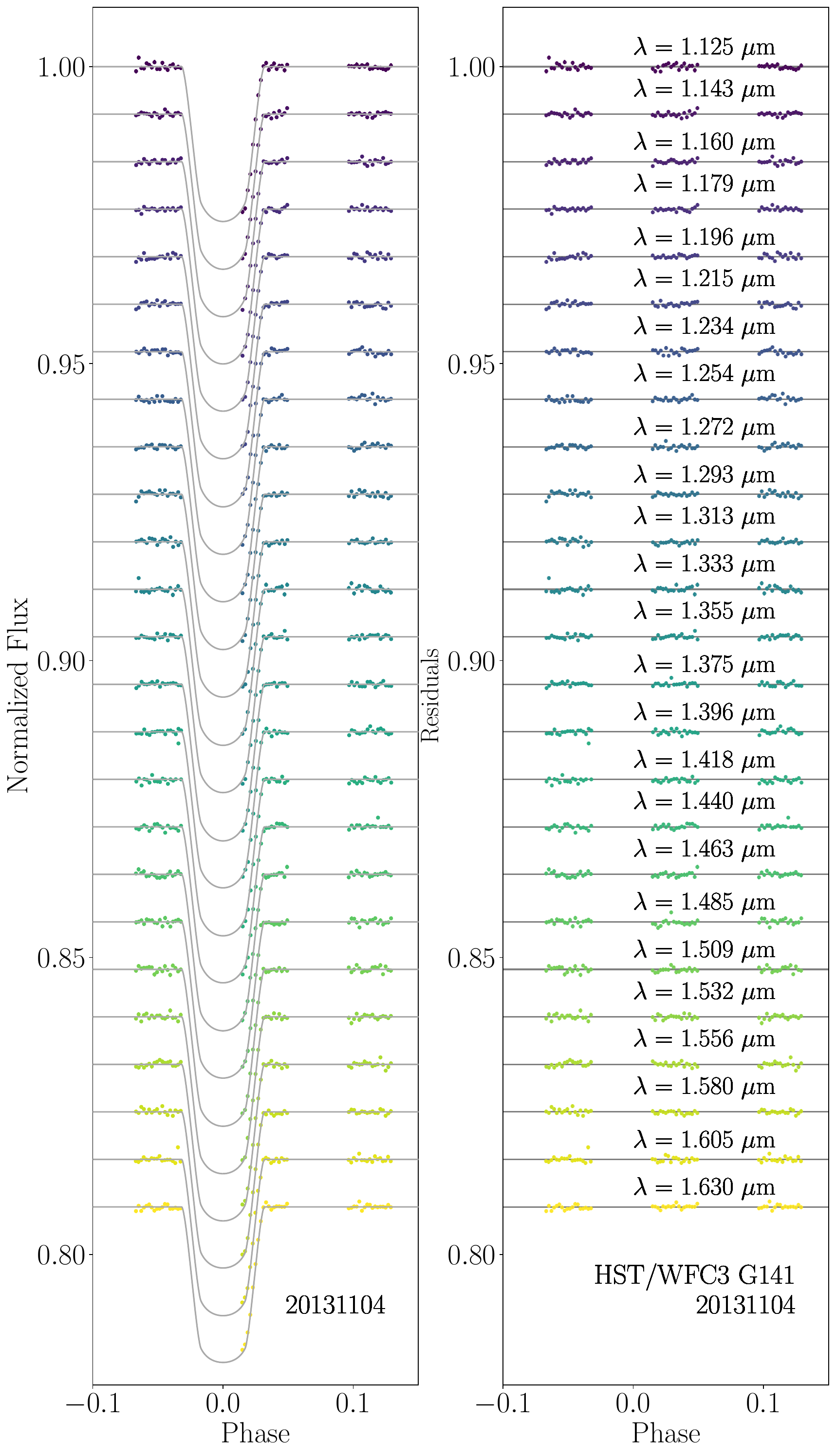} 
   \includegraphics[width=0.325\textwidth,page=1]{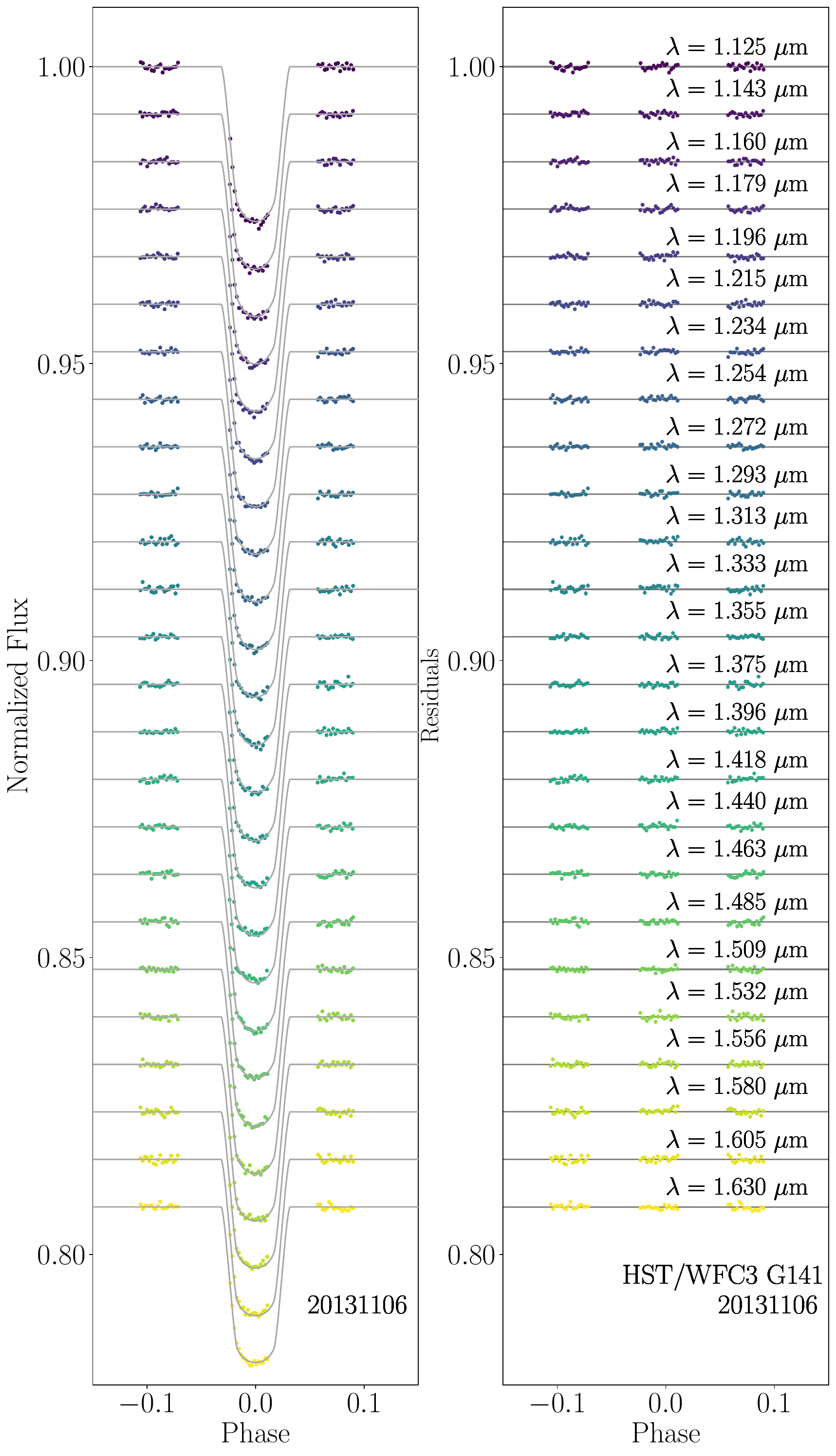} 
    \includegraphics[width=0.325\textwidth,page=1]{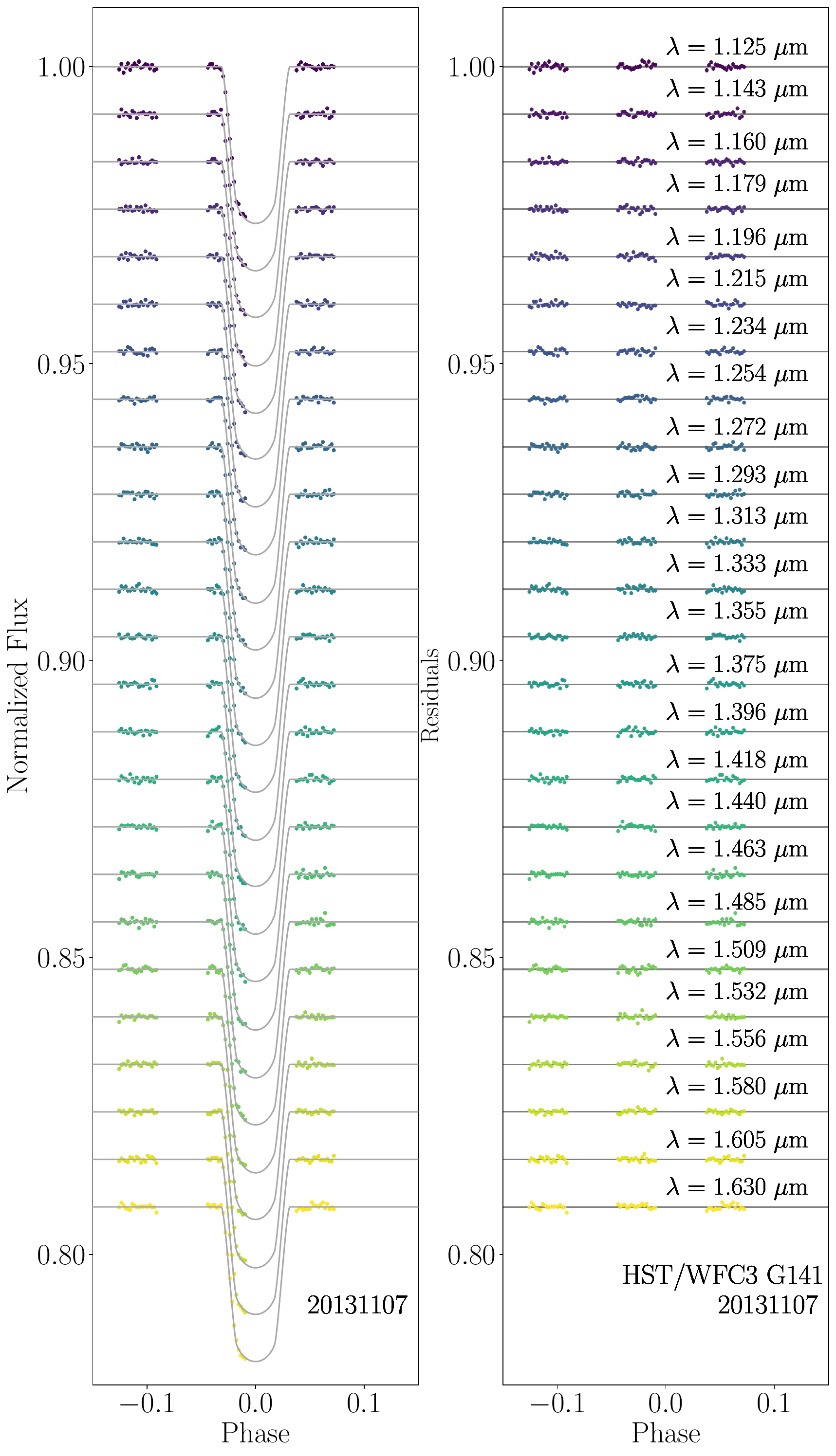} \\
    \includegraphics[width=0.325\textwidth,page=1]{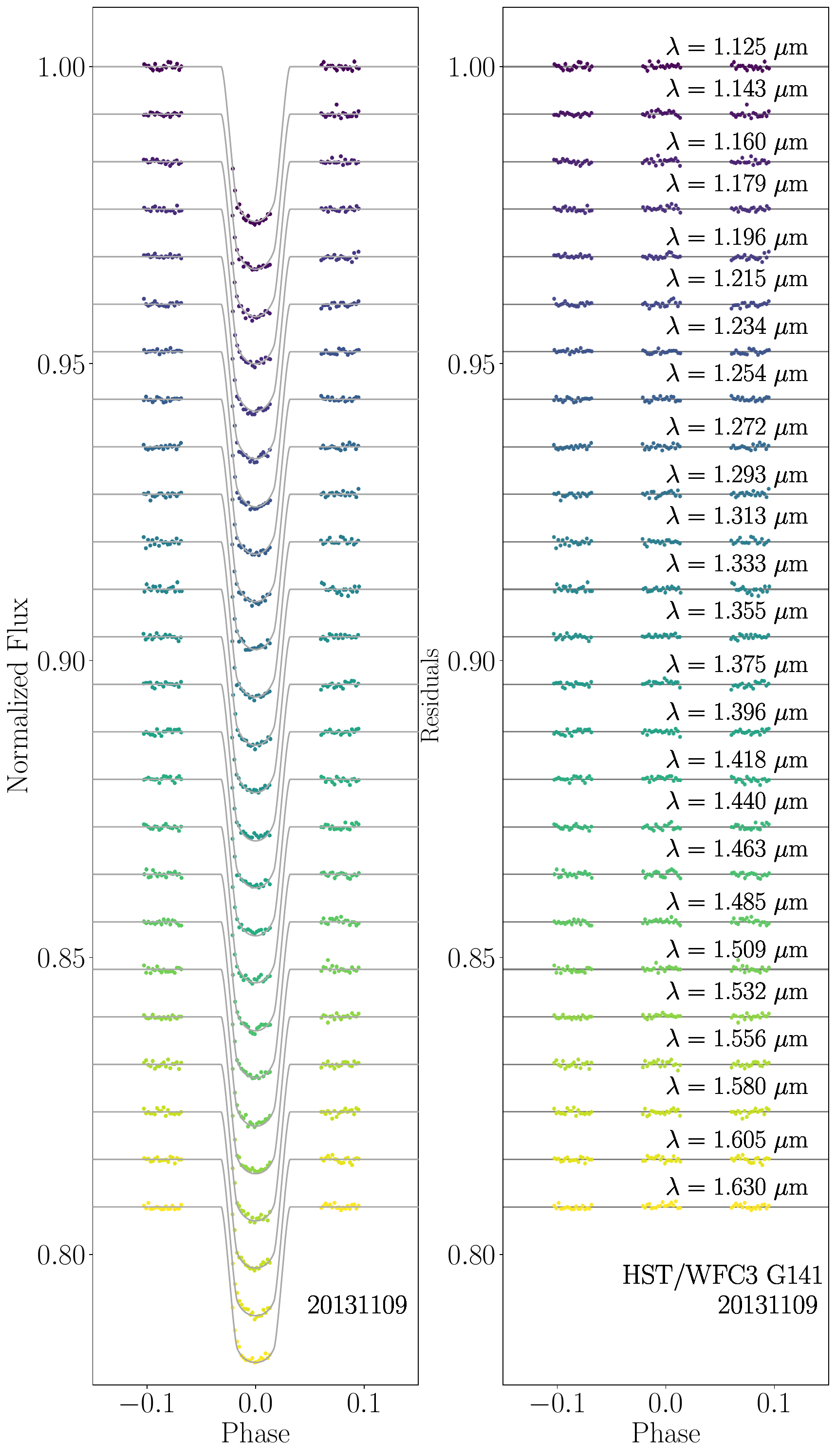} 
    \includegraphics[width=0.325\textwidth,page=1]{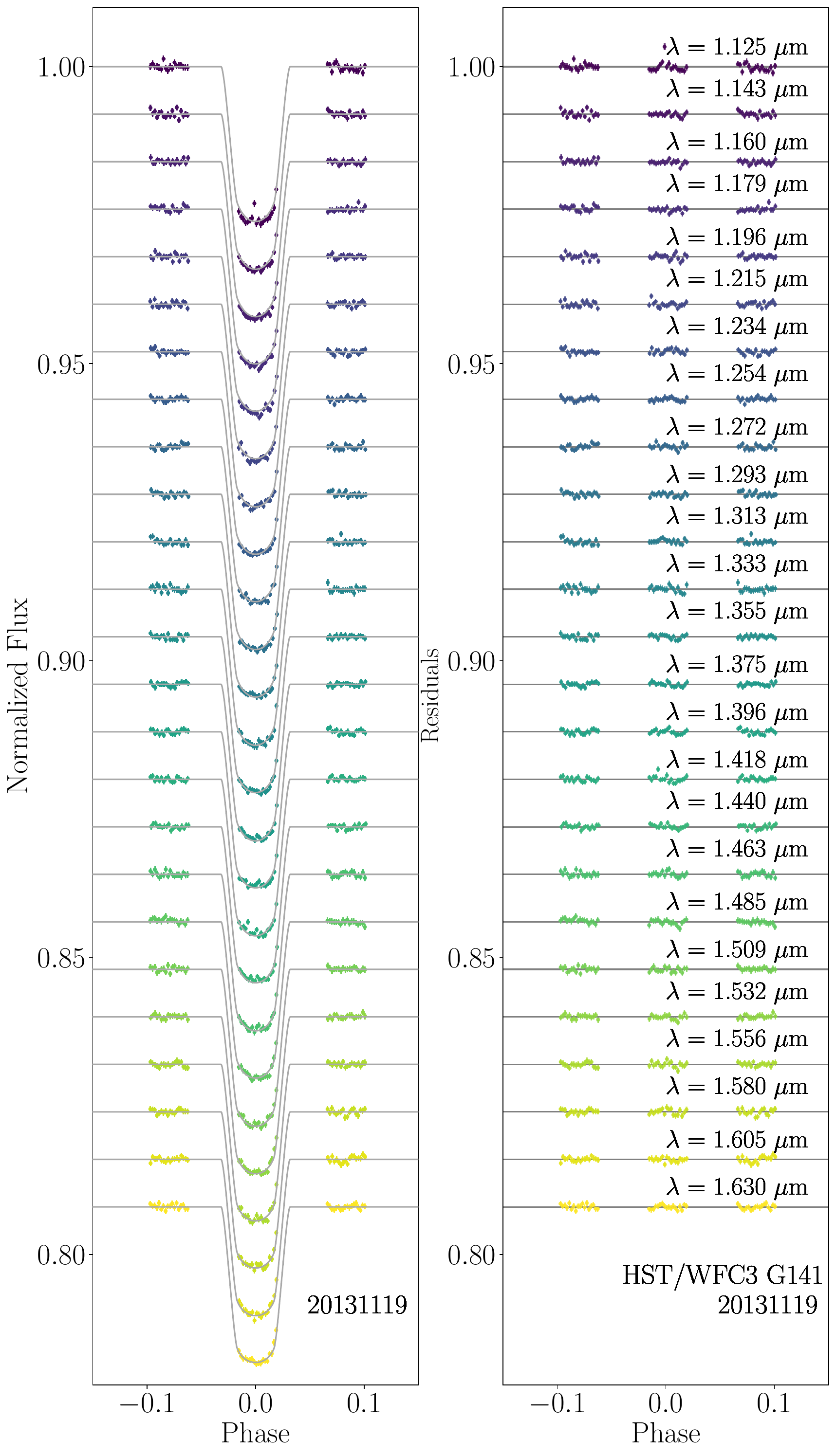} 
    \end{tabular}
    \caption{The \emph{HST} transit light curves of WASP-43~b from the observations conducted between 2013 November 04-19, respectively. The observed data are represented as dots with the \texttt{TransitFit} models are shown as solid lines. The corresponding residuals with clearly offset are displayed on the right panel.}
    \label{fig:LCs_individualHST1}
\end{figure*}

\begin{figure*}[htb]
\centering
  \begin{tabular}{cc}
    \includegraphics[width=0.5\textwidth,page=1]{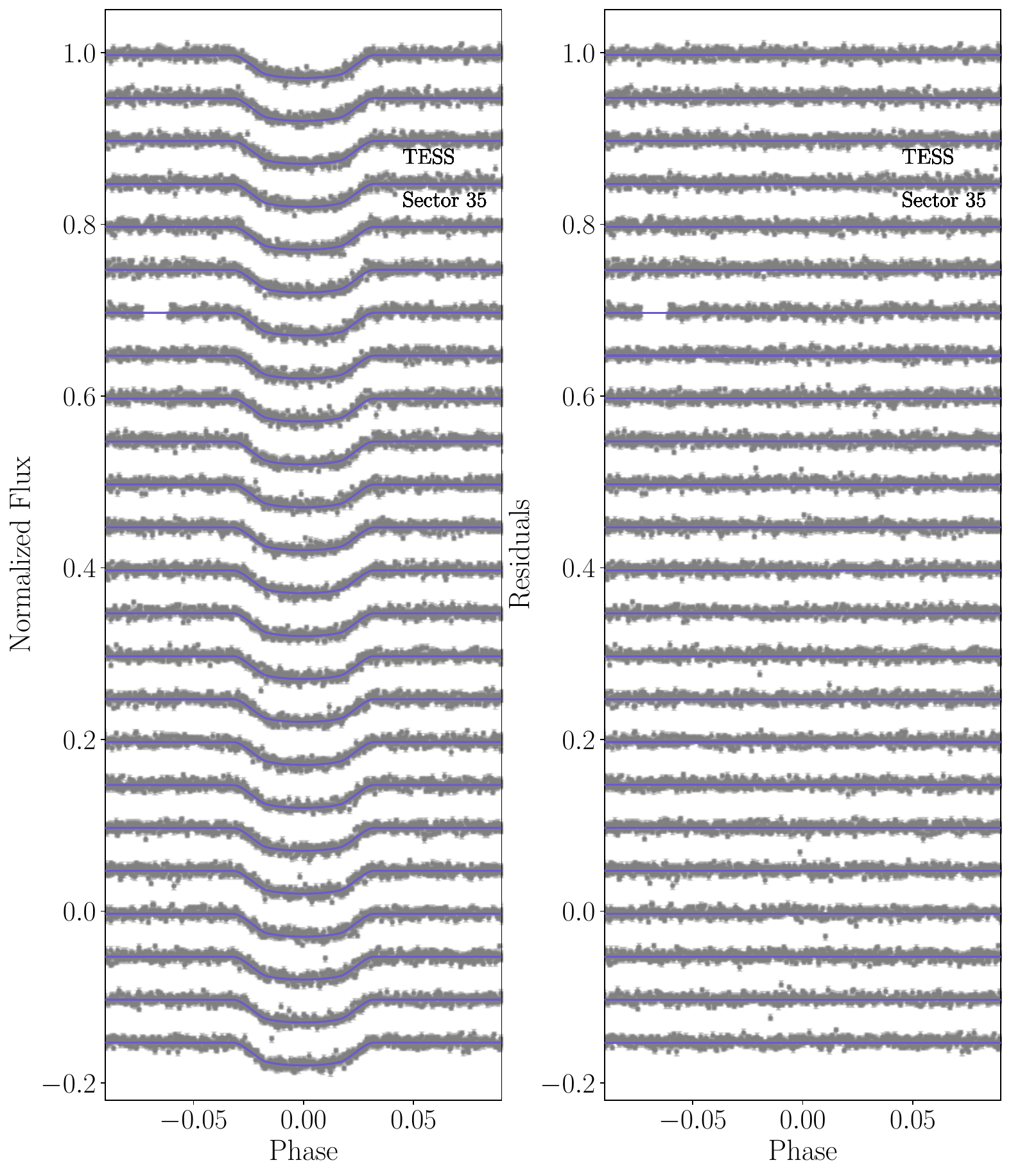} 
    \includegraphics[width=0.5\textwidth,page=1]{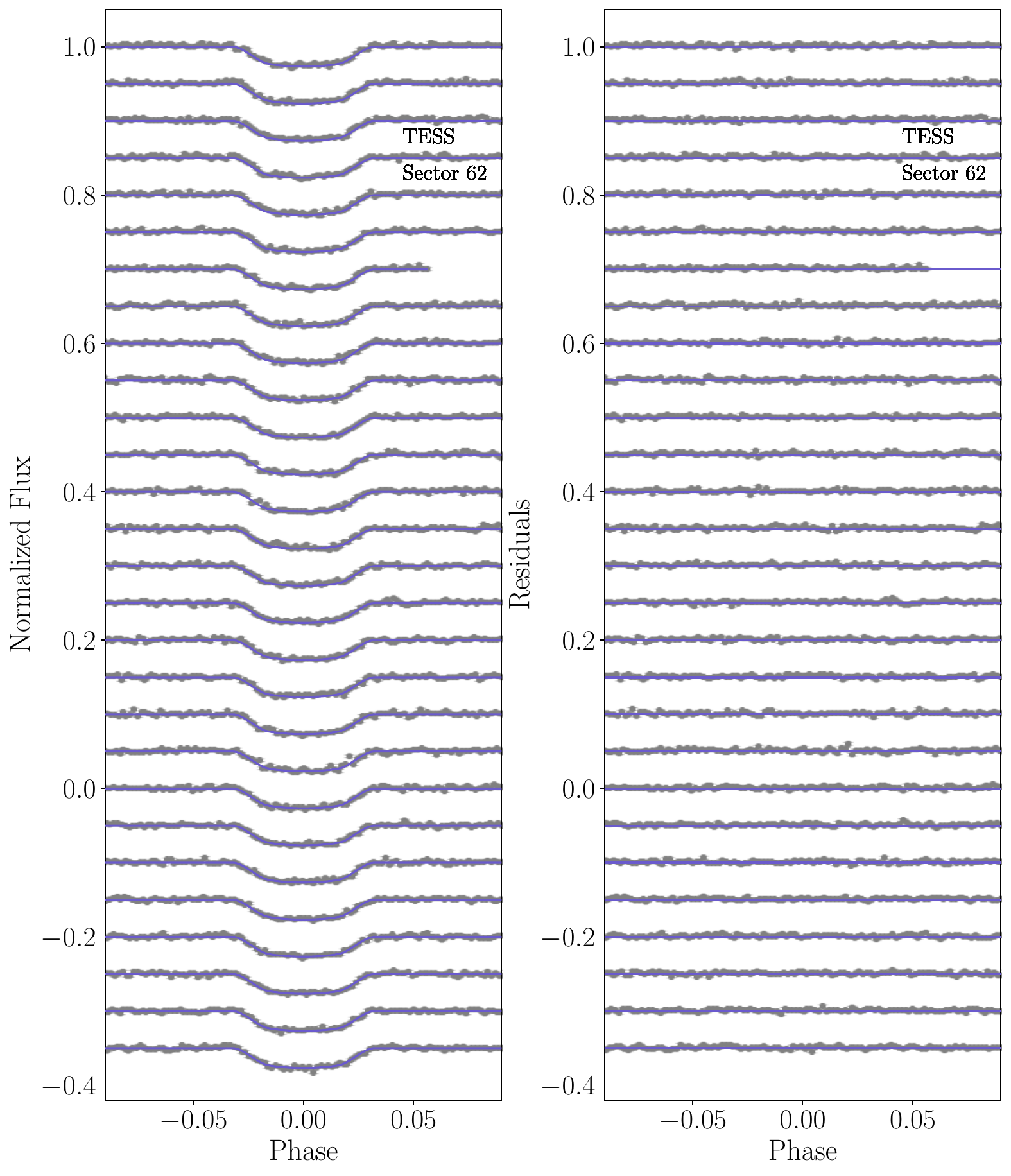} \\
    \includegraphics[width=0.5\textwidth,page=1]{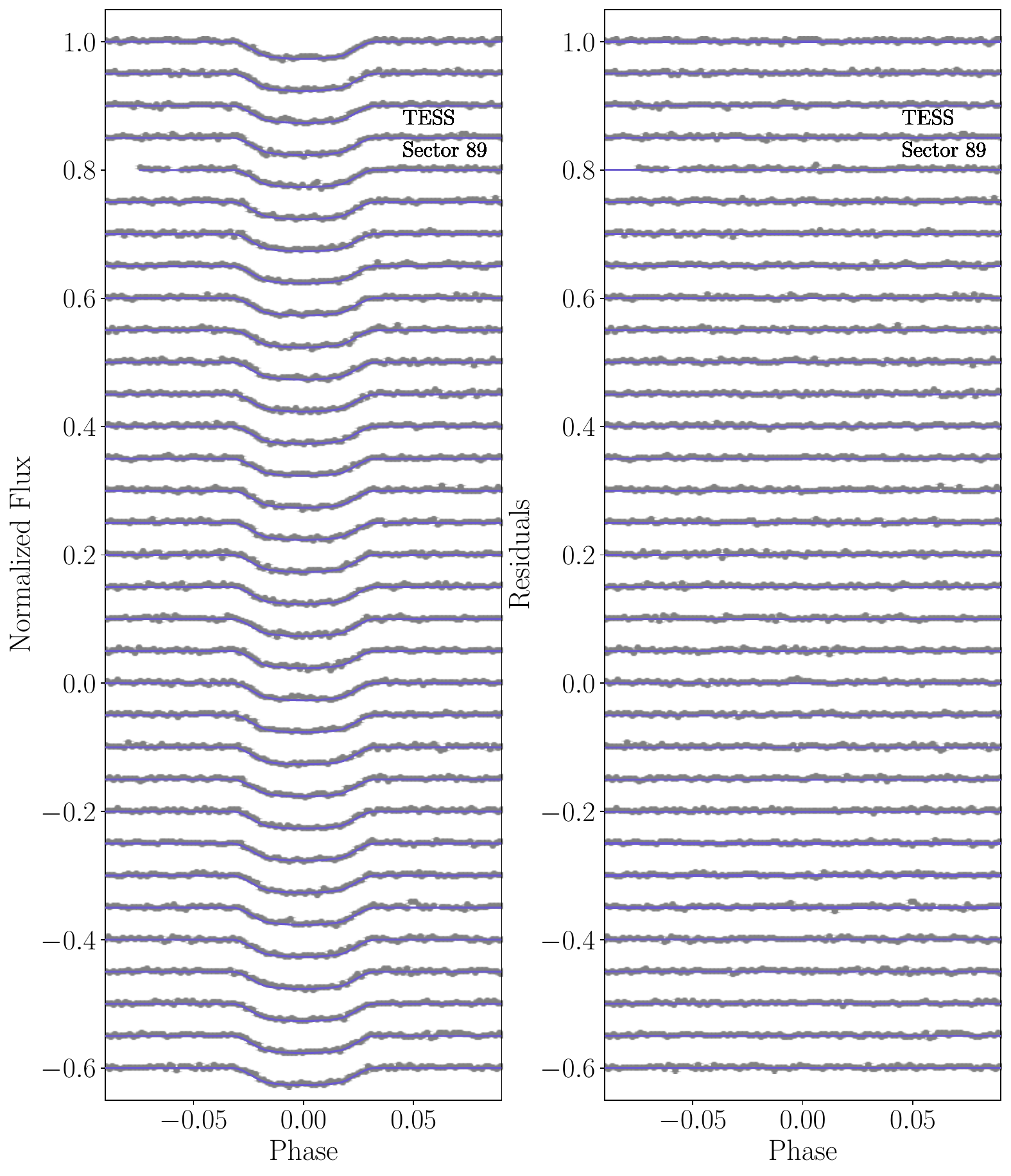}
    \end{tabular}
   \caption{The \emph{TESS} transit light curves of WASP-43~b from the observations in Sector 35, 62 and 89, respectively. The observed data are represented as dots with the \texttt{TransitFit} models are shown as solid lines. The corresponding residuals with clearly offset are displayed on the right panel.}
   \label{fig:LCs_individualTESS}
\end{figure*}

\section{Posterior probability distribution of the MCMC fitting parameters for constant-period and orbital decay models.}
\begin{figure*}[htb]
\begin{tabular}{cc}
    \includegraphics[scale=0.38]{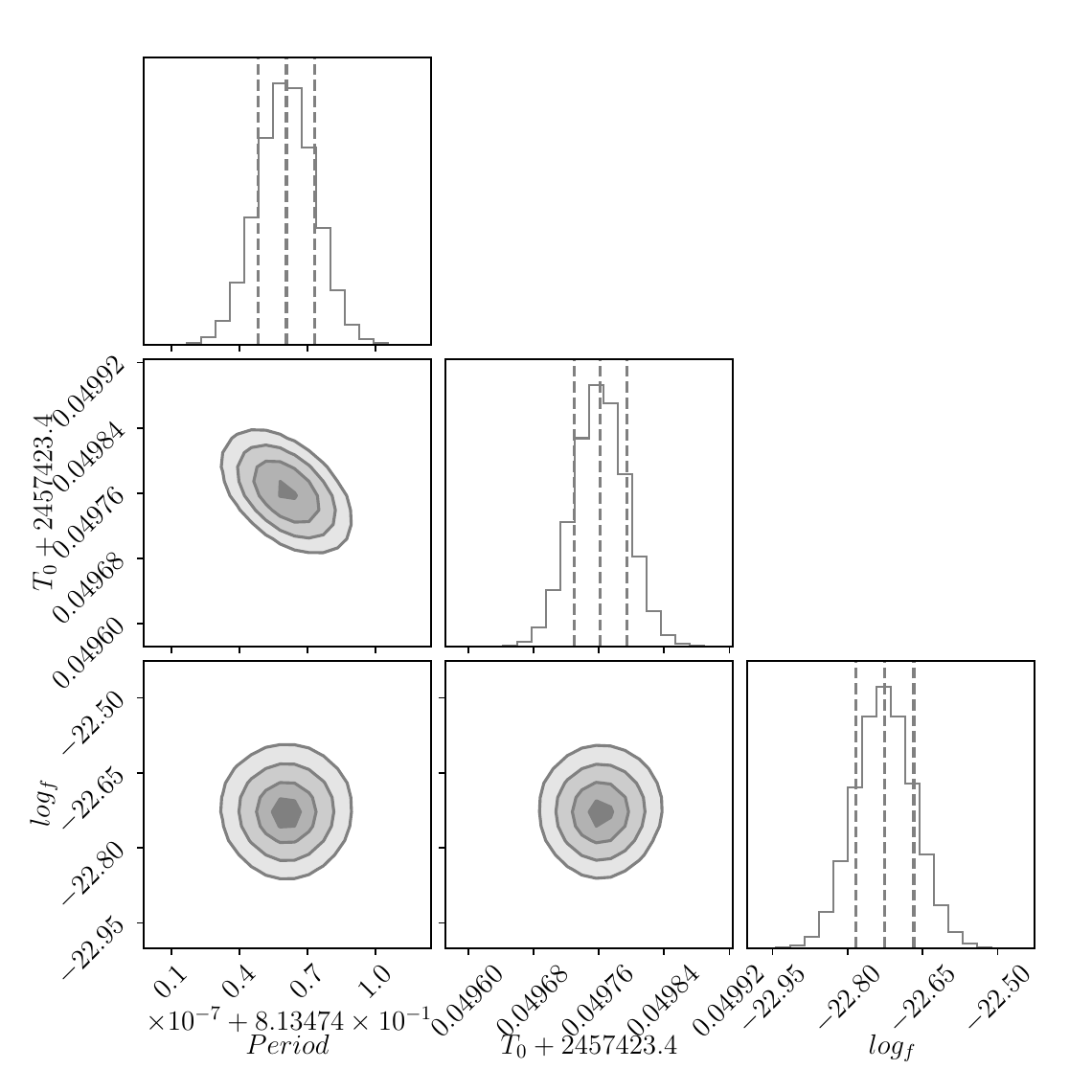} 
    \includegraphics[scale=0.4]{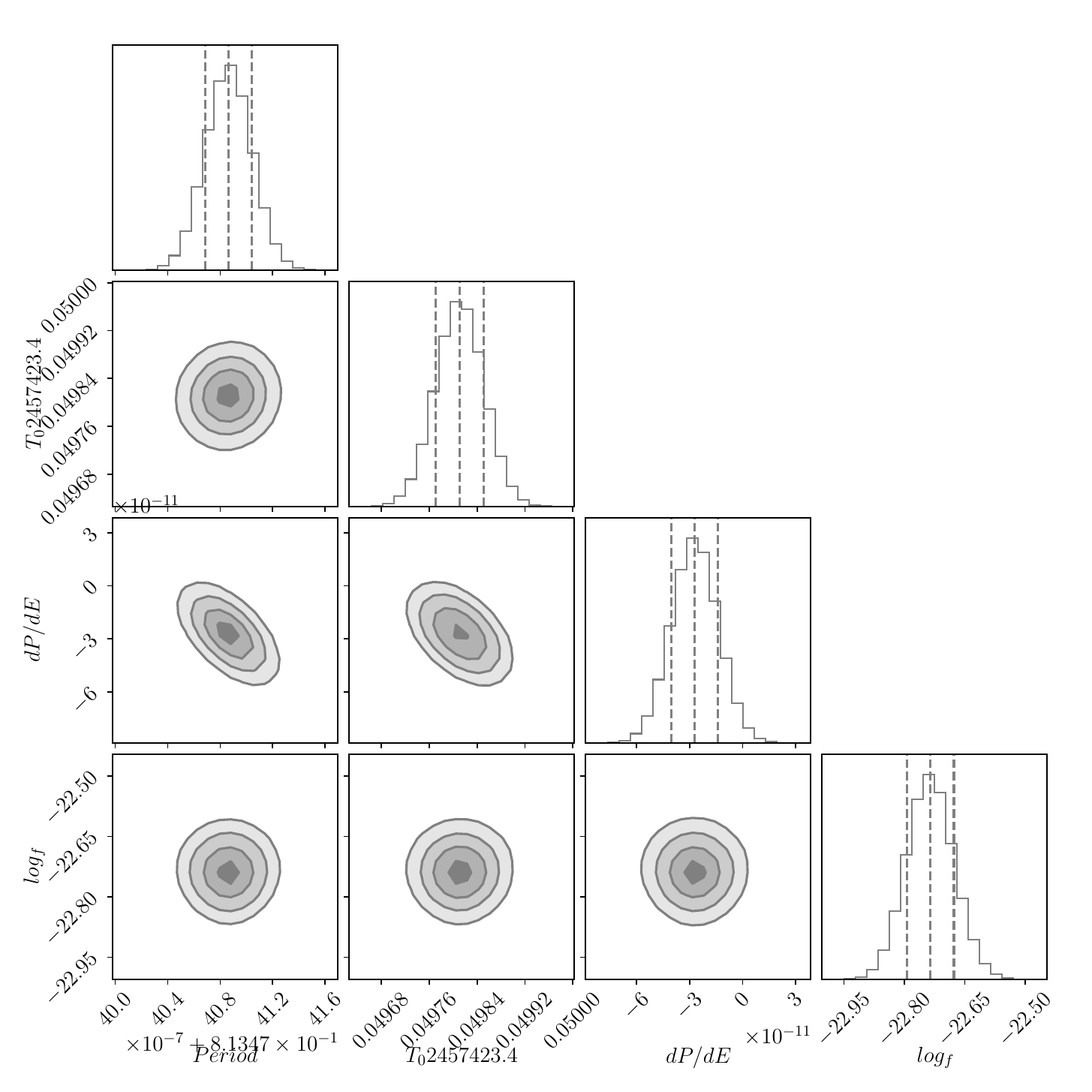} \\    
\end{tabular}
\caption{Posterior probability distribution of the constant-period and the orbital decay models MCMC fitting parameters, respectively.}
\label{fig:liDe_mcmc}
\end{figure*}

\section{Posterior probability distribution of the transmission spectrum model for WASP-43~b, using the \texttt{TauREx} package with nested sampling.}

\begin{figure*}[htb]
\centering
    \includegraphics[width=0.9\textwidth,page=1]{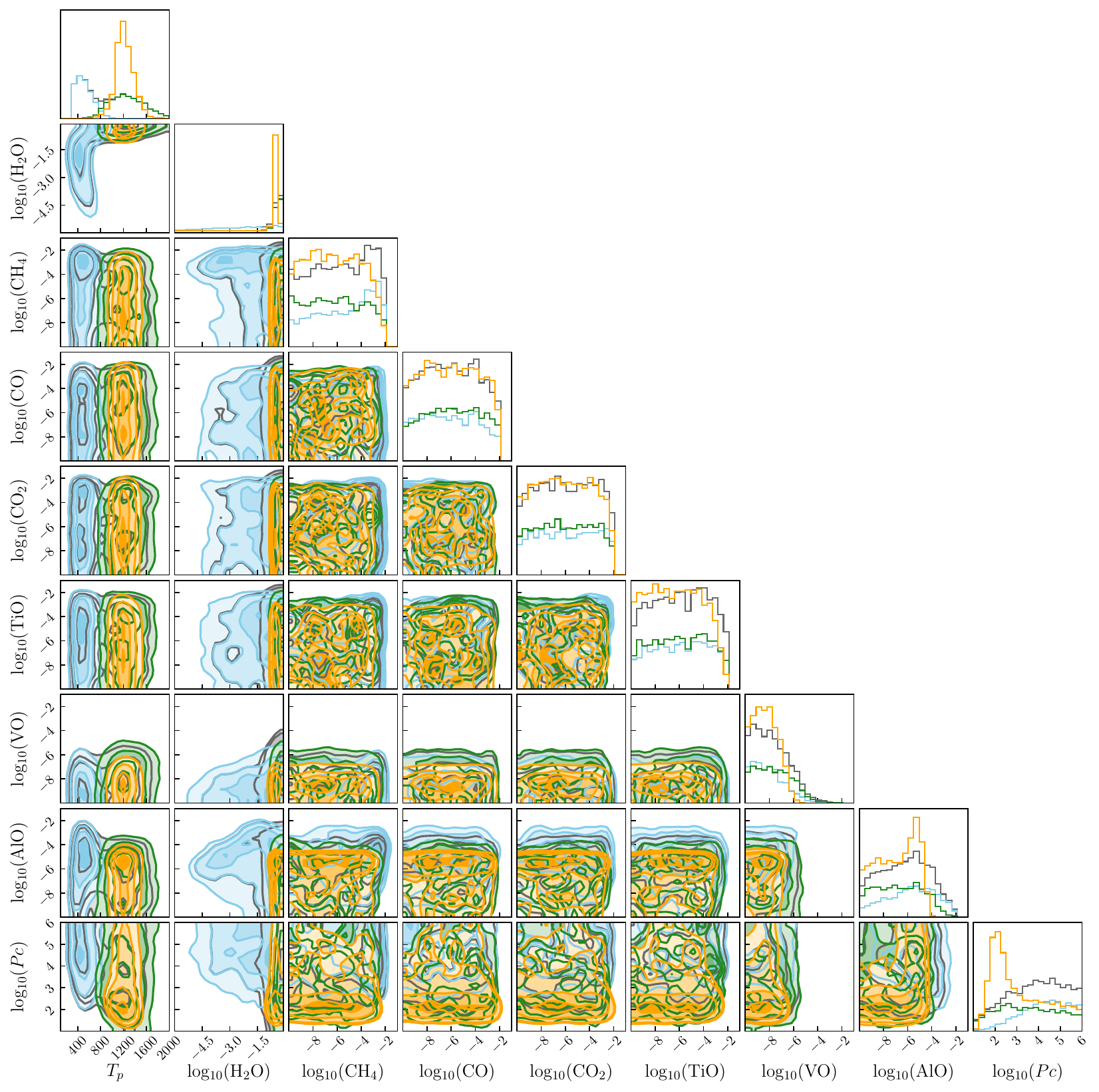} 
    \caption{Posterior probability distributions for the transmission spectrum model of WASP-43 b using only \emph{HST} data: Case I (grey), Case I.I (blue), and Case I.II (green) correspond to the \emph{HST}/\texttt{Iraclis} reductions, while Case II (orange) corresponds to the \emph{HST}/\texttt{TransitFit} reduction. The retrieved parameter values correspond to the $16^{th}$ and $84^{th}$ percentiles of the posterior distributions, representing the $1\sigma$ credible intervals.}
    \label{fig:contour-HST}
\end{figure*}
\begin{figure*}[htb]
\centering
    \includegraphics[width=0.9\textwidth,page=1]{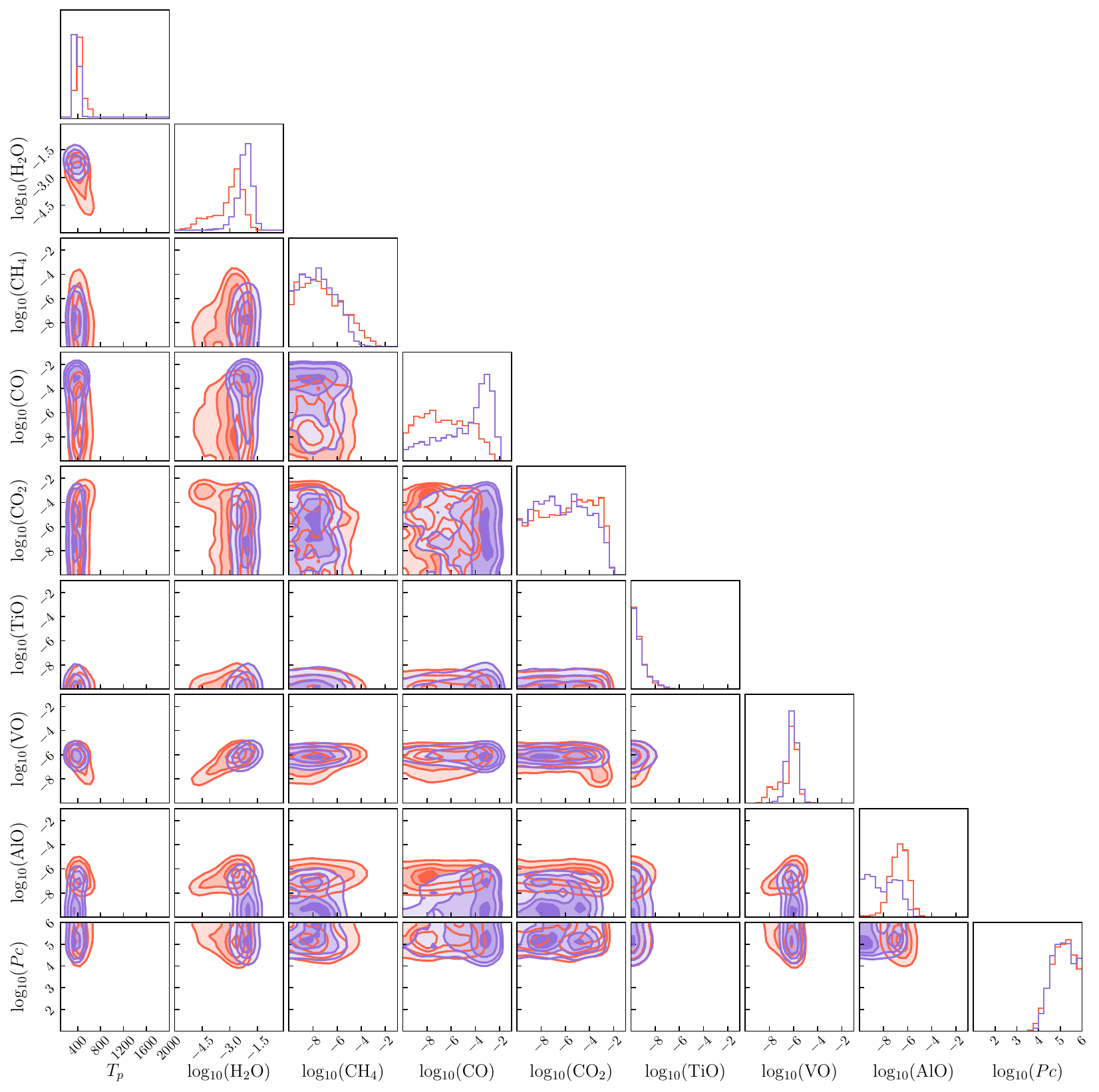} 
    \caption{Posterior probability distributions for the transmission spectrum model of WASP-43~b across the full wavelength coverage, Case: III Ground-Based+\emph{TESS}+\emph{HST}/\texttt{Iraclis}+\emph{JWST} (red) and Case: IV Ground-Based+\emph{TESS}+\emph{HST}/\texttt{TransitFit}+\emph{JWST} (purple). The retrieved parameter values correspond to the $16^{th}$ and $84^{th}$ percentiles of the posterior distributions, representing the $1\sigma$ credible intervals.}
    \label{fig:contour-GTHJ}
\end{figure*}


\bibliography{WASP43b}{}
\bibliographystyle{aasjournal}




\end{document}